\documentclass[aps,pre,twocolumn,nofootinbib,tightenlines,superscriptaddress,amsmath,amssymb,final,letterpaper]{revtex4-2}

\usepackage[utf8]{inputenc}
\usepackage{calc}
\usepackage{graphicx}
\usepackage{amsmath,amssymb,amsthm}
\usepackage{bbold}
\usepackage{bm}
\usepackage{color}
\usepackage[dvipsnames]{xcolor}
\usepackage{enumitem}
\usepackage{tikz}
\usepackage{float}
\usepackage[percent]{overpic}

\usepackage{longtable}
\usepackage{booktabs}
\usepackage{makecell}
\usepackage{multirow}
\usepackage{array}
\usepackage{threeparttable}

\DeclareMathOperator*{\argmin}{argmin}

\newcommand{\abs}[1]{\vert #1\vert}

\def\multiset#1#2{\ensuremath{\left(\kern-.3em\left(\genfrac{}{}{0pt}{}{#1}{#2}\right)\kern-.3em\right)}}

\usepackage[colorlinks=true,linkcolor=blue,urlcolor=blue,citecolor=blue,anchorcolor=blue]{hyperref}

\begin{document}
\title{Networks of amenities reveal universal homophily and heterophily across global cities}

\author{Jianrui Wu}
\affiliation{Institute of Data Science, University of Hong Kong, Hong Kong SAR, China}

\author{Baiyue He}
\affiliation{Institute of Data Science, University of Hong Kong, Hong Kong SAR, China}

\author{Alec Kirkley}
\email{alec.w.kirkley@gmail.com}
\affiliation{Institute of Data Science, University of Hong Kong, Hong Kong SAR, China}
\affiliation{Department of Urban Planning and Design, University of Hong Kong, Hong Kong SAR, China}
\affiliation{Urban Systems Institute, University of Hong Kong, Hong Kong SAR, China}

\begin{abstract}
Agglomeration economies drive urban growth at different spatial scales by enabling productivity gains, knowledge spillovers, and shared inputs among proximate firms and amenities. To develop a unified science of cities it is thus important to understand how and to what extent different amenities cluster or mix across scales and regional contexts. By utilizing a novel Bayesian framework for nonparametrically quantifying the spectrum of possible mixing patterns of amenities in a city, we identify universal spatial scales of homophily (agglomeration) and heterophily (co-agglomeration) among different amenity types across roughly 800 cities worldwide. Through a detailed longitudinal case study, we also find that the changes in heterophilic mixing derived from our methodology more effectively predict changes in neighborhood rental values than the diversity of amenities present. These findings suggest that agglomeration economies exhibit universal spatial regularities that depend largely on the types of firms or amenities being considered, rather than their specifics or regional context, and highlight the benefit of heterophilic amenity mixing at walkable spatial scales.  
\end{abstract}

\maketitle

\section{Introduction}

Agglomeration economies, defined as the benefits that firms and economic activities derive from spatial proximity, constitute a foundational concept in urban economics and economic geography. These benefits, which include enhanced productivity, innovation, and knowledge spillovers, help explain the persistence of urban concentration despite declining transportation and communication costs \citep{glaeserIntroduction2010, giulianoAgglomerationEconomiesEvolving2019}. At the industry level, spatial proximity facilitates labor pooling, input sharing, and inter-firm learning, reinforcing clustering tendencies and shaping broader metropolitan structure \citep{combesChapter5Empirics2015, carlinoChapter6Agglomeration2015, storperBuzzFacetofaceContact2004, kerrAgglomerativeForcesCluster2015, ahlfeldtEconomicsDensityEvidence2015}. Urban amenities, including restaurants, retail establishments, and personal services, represent a visible manifestation of these same forces, forming distinct concentration patterns at sub-city scales that influence neighborhood character and function \citep{leonardiAgglomerationUrbanAmenities2023, billingsAgglomerationUrbanArea2016}. Understanding the spatial organization of such amenities is therefore central to the broader scientific study of cities \citep{battyBuildingScienceCities2012}.

The mechanisms underlying agglomeration, typically classified as sharing, matching, and learning, are well established in the literature \citep{marshallPrinciplesEconomics1890, krugmanIncreasingReturnsEconomic1991, porterClustersNewEconomics1998, durantonChapter48MicroFoundations2004, glaeserCitiesAgglomerationSpatial2008, mccannTheoriesAgglomerationRegional2019}. These mechanisms operate across multiple spatial scales, and empirical evidence indicates that their externalities are strongly scale-dependent and decay rapidly, sometimes within only hundreds of meters \citep{rosenthalChapter49Evidence2004, arzaghiNetworkingMadisonAvenue2008, lavoratoriTooCloseComfort2021, hidalgoAmenityMixUrban2020}. Beyond the clustering of establishments within the same industry or category, co-agglomeration, or the spatial mixing of different amenity types, has gained increasing attention as a distinct phenomenon. Services tend to locate near suppliers and customers \citep{kolkoAgglomerationCoAgglomerationServices2007}, the local amenity mix shapes neighborhood character and livability \citep{juhaszAmenityComplexityUrban2023, chinIdentifyingUrbanFunctional2024}, and the same forces of labor pooling, input-output linkages, and knowledge spillovers drive industrial co-location as they do homogeneous clustering \citep{ellisonWhatCausesIndustry2010}. Accurately quantifying these scale-sensitive patterns therefore carries both theoretical and practical significance, with direct implications for land-use policy, zoning regulations, and place-based interventions \citep{overmanCitiesDevelopingWorld2005, neumarkChapter18PlaceBased2015, bryanCitiesDevelopingWorld2020}.

A range of metrics have been developed to characterize these spatial patterns. Area-based methods group amenities into predefined zones and evaluate their distributions using indices such as the dissimilarity index, Gini coefficient, entropy measures, or location quotients \citep{odonoghueNoteMethodsMeasuring2004, renningerUSCitiesAre2025}. These methods require the manual tuning of zone boundaries, making them highly sensitive to the modifiable areal unit problem (MAUP) \citep{openshaw1984modifiable, fotheringhamModifiableArealUnit1991}, and do not account for distances among amenities within the pre-defined regions of analysis. Point-pattern and proximity-based methods leverage exact establishment locations, overcoming these issues. For example, Ripley's $K$ function characterizes clustering across distance scales \citep{ripley1977modelling,billingsAgglomerationUrbanArea2016}, comparing spatial densities of points with those expected under a simple null model. However, this approach typically requires computationally expensive Monte Carlo simulations or approximate asymptotic results for assessing statistical significance \cite{marcon2013statistical}. Co-location quotients also compare empirical co-occurrence frequencies against random baselines \cite{leslie2011colocation,cromley2014geographically}, but have the same drawbacks. 
Perhaps most critically, these approaches often handle same-type clustering and cross-type mixing using separate statistics, null models, or parameter choices. Heterophily, or mixing preferences among amenities of different categories (also known as co-agglomeration), is therefore not typically accommodated on the same model-selection scale as homophily. Without a unified framework to quantify both homophily and heterophily on a consistent scale, systematic comparison of agglomeration tendencies across many amenity types and spatial scales remains difficult.

There is growing evidence that cities exhibit fundamental quantitative regularities despite their cultural and geographic differences. This consistency manifests across multiple domains, from predictable scaling of socioeconomic outputs and infrastructure with population size \citep{bettencourtGrowthInnovationScaling2007, bettencourtOriginsScalingCities2013, arcauteConstructingCitiesDeconstructing2015, westScaleUniversalLaws2017}, to fractal geometry of urban layouts \citep{battyFractalCitiesGeometry1994, battyNewScienceCities2013}, and common statistical rules governing human mobility and economic diversification across space \citep{gonzalezUnderstandingIndividualHuman2008, schlapferUniversalVisitationLaw2021,kirkley2020information,youn2016scaling}. Identifying these shared properties helps distinguish universal, scale-invariant laws of urbanization from local heterogeneity, a distinction essential to building a general science of cities. Whether the spatial mixing tendencies of urban amenities follow similar consistent rules across different cities remains an open question. To enable consistent large-scale comparisons across different urban contexts for revealing universality in the spatial configuration of urban amenities or points of interest, methods for extracting structural regularities in such data must be capable of capturing both homophilic and heterophilic mixing in a computationally efficient manner while avoiding tunable parameters that can artificially enforce restrictions on relevant spatial scales or enable researcher biases.

In order to evaluate the mixing tendencies of urban amenities across global cities and uncover large-scale regularities, here we view spatial agglomeration through the lens of Bayesian model selection among possible generative mechanisms underlying the proximity networks connecting amenities at different distance scales. This approach permits comparison of alternative clustering hypotheses---specifically homophilic, heterophilic, and neutral mixing patterns---using the Minimum Description Length (MDL) principle \citep{rissanen1978modeling}. The MDL principle asserts that the best model for a dataset is the one that provides the shortest description of that dataset, measured in bits of information. Previous work has applied the MDL principle to extract various structural regularities in network data, including community structure \citep{peixoto2017nonparametric,peixoto2019bayesian}, core-periphery and hub structure \citep{gallagher2021clarified,kirkley2024identifying}, node rankings \citep{peixoto2022ordered,morel2025estimation}, structural backbones \citep{kirkley2025fast,kirkley2025structural}, and clusters of similar networks \citep{coupette2022differentially,kirkley2023compressing}. MDL has also been applied to spatial data for compressing feature locations \citep{papadimitriou2005parameter}, extracting spatially cohesive regions \citep{kirkley2022spatial,morel2024bayesian}, and density estimation \citep{yang2023unsupervised,kontkanen2007mdl}. The principal advantage of MDL lies in its capacity to produce completely nonparametric data summaries that require no context-specific adaptation, facilitating straightforward comparative analyses.

In this paper we derive a unified parameter-free framework for assessing homophily and heterophily in categorical point data across space by comparing the marginal likelihoods of stochastic blockmodels \cite{peixoto2017nonparametric} that encode these various mixing patterns through proximity graphs \cite{barthelemy2011spatial}. Our analytical formulation yields exact expressions for the description lengths associated with homophilic, heterophilic and neutral (random) mixing, and the network generative process we utilize can be directly mapped to a spatial utility maximizing process for amenity location choice with agglomerative benefits that depend on proximity. Applying our framework to point-of-interest data across 779 cities worldwide, we uncover category-specific mixing regimes---homophilic or heterophilic, depending on the amenity class---that remain consistent and manifest at very similar spatial scales across diverse urban contexts. Meanwhile, longitudinal analysis of Hong Kong restaurant and rental data indicates that changes in heterophily positively predict neighborhood rent growth, while traditional amenity diversity alone is not a statistically significant predictor. These findings suggest the presence of universal mechanisms underlying agglomeration economies in global cities, and advocate for the value of co-agglomeration at walkable spatial scales. They also demonstrate that the proposed information theoretic framework provides a robust methodology for large-scale comparative urban spatial analysis, while capturing economically meaningful dimensions overlooked by conventional diversity metrics.


\section{Methods}
\label{sec:methods}

\subsection{Bayesian stochastic blockmodels (SBMs)}
\label{sec:sbms}

Since they flexibly capture both homophilic and heterophilic mixing patterns in networks (among other structural regularities \cite{young2018universality}), stochastic blockmodels (SBMs) \cite{holland1983stochastic} provide an elegant analytical framework for understanding amenity mixing using proximity networks. As we will work with the reciprocal binary notion of pairwise proximity among amenities, we focus our attention on the case of undirected, unweighted networks on $N$ nodes represented by adjacency matrices $\bm{A}$ such that $A_{ij}=1$ if nodes $i$ and $j$ have an edge and $A_{ij}=0$ otherwise. We let $E=\sum_{i<j}A_{ij}$ be the number of edges in the network and $Q=\binom{N}{2}$ be the maximum possible number of undirected edges that could have potentially formed among the $N$ nodes in the network.

Bayesian SBMs start by specifying a prior on a node partition $\bm{b}$, which assigns a label $b_i$ to each node $i$ before generating the network. A prior is also specified for a mixing matrix $\bm{p}$, which assigns a probability $p_{rs}$ for an edge to be placed between a node with label $r$ and a node with label $s$. Then, conditioned on the partition $\bm{b}$ and mixing parameters $\bm{p}$, we assess each pair of nodes $i,j$ and generate the edge $(i,j)$ with probability $p_{b_ib_j}$. It is helpful for analytical tractability and computationally efficient inference to constrain the edge probabilities $\bm{p}$ in the model by considering a symmetric variant of the SBM called the planted partition model \cite{mossel2015reconstruction}, in which there are only two independent probabilities: $p_{in}$ (for $b_i=b_j$) and $p_{out}$ (for $b_i\neq b_j$). In other words, an edge is placed between nodes $i$ and $j$ with probability $p_{in}$ if they are in the same group and $p_{out}$ if they are in different groups. In this way, we can naturally model homophilic mixing patterns using a planted partition model with $p_{in}>p_{out}$ and heterophilic mixing patterns with $p_{out}>p_{in}$. The case $p_{in}=p_{out}=p$ corresponds to a standard Erd\H{o}s--R\'enyi random graph in which all edges are placed with a constant probability $p$.

A natural prior on the node partition $\bm{b}$ can be constructed from the following generative process: (1) generate the number of groups $B$ uniformly at random in $1,...,N$; (2) generate the $B$ groups' sizes $\bm{n}=\{n_1,...,n_B\}$ uniformly at random, where $n_r=\sum_{i=1}^{N}\delta_{b_i,r}$ is the number of nodes with label $r$ ($\delta_{xy}$ is the Kronecker delta function which is $1$ if and only if $x=y$ and $0$ otherwise); (3) generate the partition $\bm{b}$ uniformly at random given the constraints imposed by $\bm{n}$ and $B$. The full prior on the partition $\bm{b}$ can then be written as
\begin{align}\label{eq:prior-b}
P_{part}(\bm{b}) = \frac{1}{N}\times \frac{1}{{N-1\choose B-1}}\times \frac{1}{{N\choose n_1,...,n_B}},  \end{align}
where ${n\choose k}$ is the binomial coefficient and ${n\choose k_1,...,k_B}$ is the multinomial coefficient. This hierarchical uniform prior allows for an assumption of indifference on the structure of the partition $\bm{b}$ while avoiding the strong bias towards very small groups that results from a completely uniform prior \cite{peixoto2019bayesian}. We also have a prior $P_{mix}(p_{in},p_{out})$ on the mixing parameters, which we will discuss further in Sec.~\ref{sec:proximity-sbms}.

Under the planted partition model the likelihood $P_{ppm}(\bm{A}\vert \bm{b},p_{in},p_{out})$ of observing a network $\bm{A}$ given the node partition $\bm{b}$ and mixing probabilities $p_{in},p_{out}$ is
\begin{align}\label{eq:ppm-likelihood}
P_{ppm}&(\bm{A}\vert \bm{b},p_{in},p_{out}) = \prod_{i<j}p_{b_ib_j}^{A_{ij}}(1-p_{b_ib_j})^{1-A_{ij}}\\
&=p_{in}^{E_{in}}(1-p_{in})^{Q_{in}-E_{in}}p_{out}^{E_{out}}(1-p_{out})^{Q_{out}-E_{out}},
\end{align}
where $E_{in} = \sum_{i<j}A_{ij}\delta_{b_ib_j}$, $E_{out} = \sum_{i<j}A_{ij}(1-\delta_{b_ib_j})$, $Q_{in} = \sum_{i<j}\delta_{b_ib_j}$, and $Q_{out} = \sum_{i<j}(1-\delta_{b_ib_j})$ are the number of within- and between-group edges observed in $\bm{A}$ and the maximum possible number of within- and between-group edges given the node partition $\bm{b}$, respectively. Note that $E_{in}+E_{out}=E$ is the total number of edges in $\bm{A}$ and $Q_{in}+Q_{out}=Q=\binom{N}{2}$ is the maximum possible number of edges in $\bm{A}$.

In most applications of SBMs, the optimal partition $\bm{b}$ and mixing parameters $\bm{p}$ corresponding to a given observed network $\bm{A}$ are then inferred using maximum a posteriori (MAP) estimation by maximizing the product of the likelihood and priors over $\bm{b},\bm{p}$. It is also common to sample these node partitions and/or mixing parameters from the appropriate posterior distribution for a fully Bayesian approach with uncertainty quantification \cite{peixoto2019bayesian}.

\subsection{Models of amenity mixing}
\label{sec:proximity-sbms}

We can now apply the SBM framework to understand the mixing structure of spatial amenities while nonparametrically selecting for the best description of the data as homophilic, heterophilic, or neutral. We are given a dataset of $N$ amenities $i=1,2,...,N$, each of which is associated with a spatial location $\bm{x}_i$ and a label $b_i$ that categorizes the amenity $i$ into one of $B>1$ disjoint classes. We denote the vector of coordinate pairs for all points with the matrix $\bm{X}=\{\bm{x}_i\}_{i=1}^{N}$, and the vector of all amenity labels as $\bm{b}=\{b_1,...,b_N\}$. Our starting point is to determine, using the MDL principle \cite{rissanen1978modeling}, whether or not the point configuration specified by $\{\bm{X},\bm{b}\}$ is best described as homophilic, heterophilic, or neutral (neither homophilic nor heterophilic). To operationalize these concepts in an analytically tractable way, we will consider a purely relational point of view and construct proximity networks among the amenities, which can be modeled using a specially designed variant of the Bayesian SBM formulation of Sec.~\ref{sec:sbms}.

Going back to the prior $P_{mix}(p_{in},p_{out})$ in the Bayesian SBM, we can define mechanisms that generate homophilic (abbreviated ``hom'' in the model specification), heterophilic (``het''), and neutral mixing (``neu'') while permitting nonparametric inference through marginalization of the model likelihood. As stated before, in the planted partition model $p_{in}>p_{out}$ generates homophilic networks, $p_{in}<p_{out}$ generates heterophilic networks, and $p_{in}=p_{out}=p$ generates purely random graphs at some density $p$. To express indifference on the density of mixing in the network while preserving the desired overall mixing structure (hom, het, neu), we can construct the following three priors on the mixing probabilities
\begin{align}
\label{eq:prior-hom}
P^{(hom)}_{mix}(p_{in},p_{out}) &= 2\Theta(p_{in}-p_{out}),\\ 
\label{eq:prior-het}
P^{(het)}_{mix}(p_{in},p_{out}) &= 2\Theta(p_{out}-p_{in}),\\ 
\label{eq:prior-neu}
P^{(neu)}_{mix}(p_{in},p_{out}) &= \delta(p_{in}-p_{out}),
\end{align}
where $\delta(x-y)$ is the Dirac delta function, $\Theta(x-y)$ is the Heaviside step function such that $\Theta(x-y)=1$ iff $x\geq y$ and $0$ otherwise, and the factor of $2$ is required for normalization of the homophilic and heterophilic priors.

We can then compute the marginal likelihood of the homophilic or heterophilic planted partition specifications using
\begin{align}\label{eq:marginal-lik}
P_{ml}(\bm{A},\bm{b}) &= \int_{0}^{1}\int_{0}^{1} P_{ppm}(\bm{A}\vert \bm{b},p_{in},p_{out})\\
&~~~~~~~~~~~~~\times P_{mix}(p_{in},p_{out})P_{part}(\bm{b})dp_{in}dp_{out}, \nonumber
\end{align}
where $P_{mix}$ is $P^{(hom)}_{mix}$ for the homophilic model and $P^{(het)}_{mix}$ for the heterophilic model. The marginal likelihood $P^{(hom)}_{ml}(\bm{A},\bm{b})$ corresponds to the probability that an observed network $\bm{A}$ and node partition $\bm{b}$ were generated using the homophilic planted partition model, and $P^{(het)}_{ml}(\bm{A},\bm{b})$ is the probability that they were generated using the heterophilic planted partition model. We can compute the marginal likelihood $P^{(neu)}_{ml}(\bm{A})$ of the neutral model specification in the same way but can omit $P(\bm{b})$---since the edge probabilities are uniform, the group labels are not used in the model and are ignored in the marginal likelihood. In the neutral case, $P^{(neu)}_{ml}(\bm{A})$ is the probability of generating a network $\bm{A}$ from an Erd\H{o}s--R\'enyi random graph on $N$ nodes with the number of edges $E$ chosen uniformly at random. 

\begin{figure*}
	\centering
	\includegraphics[width=\linewidth]{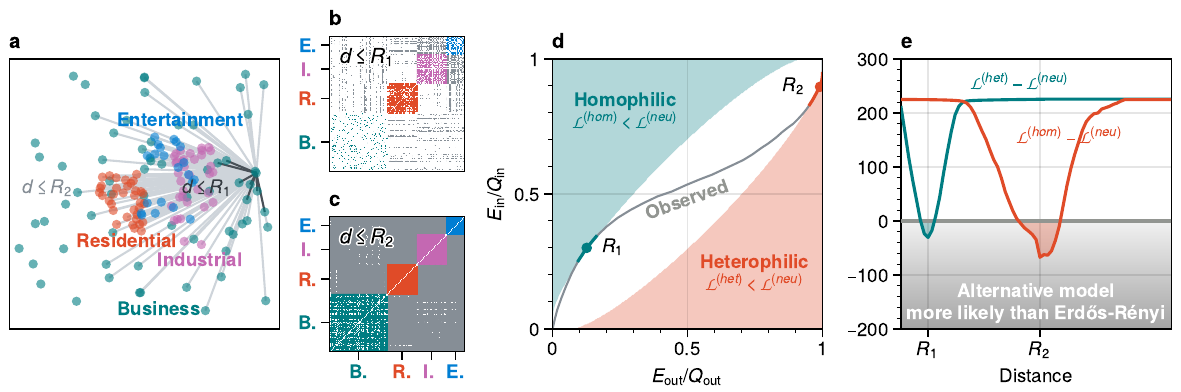}
	\caption{
		\textbf{Bayesian nonparametric model of amenity mixing.} (a) Example amenity layout with four (synthetic) amenity types, showing the edges of the proximity networks $\bm{A}(\bm{X},\epsilon)$ at small ($\epsilon=R_1$) and large ($\epsilon=R_2$) distance thresholds. (b,c) Adjacency matrices for these proximity networks, exhibiting homophilic mixing at $\epsilon=R_1$ and heterophilic mixing at $\epsilon=R_2$. (d) Model selection phase diagram showing the trajectory of the example system as we increase the distance scale $\epsilon$ in $[R_1,R_2]$. The white region indicates the spatial scale regime in which neither mixing model is more parsimonious than the model of random amenity mixing. (e) Description lengths of homophilic and heterophilic models (Eqs.~\ref{eq:dl-hom}~and~\ref{eq:dl-het}), subtracting off the baseline description length from the random mixing model (Eq.~\ref{eq:dl-neu}). These curves highlight the same spatial regimes of homophily and heterophily highlighted in the phase diagram.    
	}
	\label{fig:diagram}
\end{figure*}

These marginal likelihoods allow us to assign a description length to the data $\bm{A},\bm{b}$ using Shannon's source coding theorem \cite{shannon1948mathematical}, which states that the shortest lossless binary encoding of objects sampled from a distribution $P(x)$ will assign codelength $-\log P(x)$ to object $x$, where we use $\log \equiv \log_2$ for brevity. Thus, the homophilic, heterophilic, and neutral models have description lengths $\mathcal{L}_{hom}=-\log P^{(hom)}_{ml}(\bm{A},\bm{b})$, $\mathcal{L}_{het}=-\log P^{(het)}_{ml}(\bm{A},\bm{b})$ and $\mathcal{L}_{neu}=-\log P^{(neu)}_{ml}(\bm{A})$ respectively. For an observed network $\bm{A}$ with node labels $\bm{b}$, the three mixing hypotheses can then be compared on the basis of how much they compress the data, using $\mathcal{L}_{hom},\mathcal{L}_{het},\mathcal{L}_{neu}$. The model with the shortest description length is the best description of the data according to the MDL principle, providing a principled nonparametric criterion for determining whether a given set of pairwise relationships exhibits homophily, heterophily, or neither.

SBMs were initially developed as a generative model to capture homophily in social networks \cite{holland1983stochastic}. The concepts of homophily and heterophily in networks have also been generalized to accommodate scalar numerical attributes, most commonly the degrees of network nodes, and in this context homophily and heterophily are called assortativity and disassortativity respectively \cite{newman2003mixing}. Meanwhile, the terms agglomeration and co-agglomeration are most commonly used in the economics literature to describe homophilic and heterophilic tendencies in the co-location of firms and amenities across space  \cite{kolkoAgglomerationCoAgglomerationServices2007}.  In this paper, we use the terms homophily (heterophily), assortativity (disassortativity), and agglomeration (co-agglomeration) interchangeably, with a preference for homophily and heterophily as they are perhaps the most widely used across applications.

We can now apply our modified SBM variant to spatial amenities by first defining the proximity network $\bm{A}(\bm{X},\epsilon)$ for the dataset $\bm{X}$ at distance scale $\epsilon$ as the adjacency matrix with entries
\begin{align}
A_{ij}(\bm{X},\epsilon) = \begin{cases}
1,~d(\bm{x}_i,\bm{x}_j) \leq \epsilon \\
0,~\text{otherwise},
\end{cases}    
\end{align}
where the distance $d(\bm{x}_i,\bm{x}_j)$ is most simply taken to be Euclidean, but can in principle be modified to account for the earth's curvature or constraints imposed by transportation networks \cite{poudyal2023characterizing}. We denote the total number of edges in $\bm{A}(\bm{X},\epsilon)$ as $E(\bm{X},\epsilon)=\sum_{i<j}A_{ij}(\bm{X},\epsilon)$. By varying the distance scale $\epsilon$, the proximity network $\bm{A}(\bm{X},\epsilon)$ allows us to understand the pairwise distance relationships among the points in $\bm{X}$ at arbitrary scales. By considering a network of pairwise proximity among the points in space, we focus on the relational structure intrinsic to homophily and heterophily rather than the exact spatial locations of the points. This relational point of view permits simple analytical treatment through SBMs, without requiring the asymptotic approximations used for spatial processes \cite{marcon2013statistical}.

Now, using the marginal likelihood integrals, we can substitute in the proximity graph $\bm{A}(\bm{X},\epsilon)$ to construct description lengths for comparing homophilic, heterophilic, and neutral hypotheses of spatial mixing among urban amenities:
\begin{align}
\label{eq:dl-hom}
\mathcal{L}_{\bm{X},\bm{b}}^{(hom)}(\epsilon) &= -\log P^{(hom)}_{ml}(\bm{A}(\bm{X},\epsilon),\bm{b})\\
\label{eq:dl-het}
\mathcal{L}_{\bm{X},\bm{b}}^{(het)}(\epsilon) &= -\log P^{(het)}_{ml}(\bm{A}(\bm{X},\epsilon),\bm{b})\\
\label{eq:dl-neu}
\mathcal{L}_{\bm{X},\bm{b}}^{(neu)}(\epsilon) &= -\log P^{(neu)}_{ml}(\bm{A}(\bm{X},\epsilon)).
\end{align}
In Appendix~\ref{sec:dl-derivations} we compute exact expressions for these three description lengths, and in Appendix~\ref{sec:algorithm} we describe an efficient algorithm for computing them numerically given an amenity dataset $\bm{X},\bm{b}$. In Appendix~\ref{sec:analytical-properties}, we analytically derive a phase transition with respect to the density of edges between amenities of different types, which separates the homophilic and heterophilic regimes and provides additional intuition as to the behavior of the amenity classification method in practice.

It is worth noting that the ensemble of possible proximity networks among points in space is usually more constrained than the full SBM ensemble due to the spatial nature of the dataset, since the points must satisfy the triangle inequality if $d$ is a metric. However, for analytical tractability we do not account for this transitivity here, allowing $d$ to in principle be a non-metric distance and the standard marginal likelihoods $P_{ml}$ to be used for the proximity networks $\bm{A}(\bm{X},\epsilon)$. This also allows us to focus on the purely relational aspect of pairwise co-location among amenities rather than proximity induced by spatial constraints. However, in Appendix~\ref{sec:physical} we show that the SBM generative process for a proximity network among amenities directly corresponds to the marginal likelihood of this proximity network in a spatial location choice model where agglomerative benefits depend on both distance and amenity type. This links our SBM to a simple hypothesis of a co-location mechanism among different types of amenities in space.

Using this MDL model selection framework, we can determine the regimes (i.e., ranges of the distance scale $\epsilon$) in which different types of amenities exhibit different mixing patterns. We can also compute coarser summary statistics to more succinctly capture the mixing patterns within a particular urban area for large scale comparative analyses. One meaningful measure we can extract from our framework is the characteristic spatial scale associated with each mixing classification, given by
\begin{align}\label{eq:scale}
\epsilon_{hom/het}^{\ast} = \argmin_{\epsilon} \Big\{\mathcal{L}^{(hom/het)}(\epsilon)\Big\}.
\end{align}
This captures the scale at which the model of interest most parsimoniously summarizes the mixing structure of the amenities. $\epsilon_{hom/het}^{\ast}$ is primarily meaningful when $\mathcal{L}^{(hom/het)}(\epsilon)-\mathcal{L}^{(neu)}(\epsilon)<0$, as this indicates that the associated model better compressed the data than the model of random mixing. Another measure we can compute based on the 
description length trajectories is the compression ratio
\begin{align}\label{eq:ratio}
\eta^{(hom/het)}(\epsilon) = \frac{\mathcal{L}^{(neu)}(\epsilon)}{\mathcal{L}^{(hom/het)}(\epsilon)},
\end{align}
which characterizes how the structured mixing model (homophilic or heterophilic) compresses relative to the random mixing model at the scale $\epsilon$. $\eta > 1$ indicates that the structured model compresses better, while $\eta<1$ indicates that the data are more parsimoniously described as mixing randomly in space. These measures are explored along with the raw description length values in Sec.~\ref{sec:results}.

In Fig.~\ref{fig:diagram} we show a diagram of the proposed method, using a synthetic amenity layout with $B=4$ distinct amenity types (Entertainment, Business, Residential, Industrial, which are labeled just for illustration purposes). In Fig.~\ref{fig:diagram}(a) we plot the configuration of amenities in space, with Entertainment, Residential, and Industrial amenities exhibiting clear agglomerative behavior, and Business amenities exhibiting a more dispersed spatial configuration. We also show two distance scales $\epsilon=R_1$ and $\epsilon=R_2$ at which proximity networks are constructed. In Fig.~\ref{fig:diagram}(b) and Fig.~\ref{fig:diagram}(c) we plot the adjacency matrices $\bm{A}(\bm{X},R_1)$ and $\bm{A}(\bm{X},R_2)$ corresponding to these two distance scales, coloring within-group edges by their amenity color and between-group edges in gray. We can see a clear densification of between-group edges at larger distance scales, while homophilic mixing dominates at smaller scales. In Fig.~\ref{fig:diagram}(d) we plot a phase diagram of the homophilic, heterophilic, and neutral regimes as a function of the in-edge and out-edge densities $\hat p_{in},\hat p_{out}$, finding a transition from homophilic to heterophilic mixing as we move from $R_1$ to $R_2$. The white region is the neutral zone for which the model of random mixing is optimal in terms of description length, since it is too costly (in terms of bits of information) to transmit the node labels $\bm{b}$ when there is little amenity mixing structure that can be exploited for further data compression. Finally, in Fig.~\ref{fig:diagram}(e) we show the differences $\mathcal{L}^{(hom)}_{\bm{X},\bm{b}}(\epsilon)-\mathcal{L}^{(neu)}_{\bm{X},\bm{b}}(\epsilon)$ and $\mathcal{L}^{(het)}_{\bm{X},\bm{b}}(\epsilon)-\mathcal{L}^{(neu)}_{\bm{X},\bm{b}}(\epsilon)$ between the homophilic and heterophilic mixing models and the neutral model, as a function of distance scale $\epsilon$ for the example dataset in Fig.~\ref{fig:diagram}(a). We find that at distance scale $\epsilon \sim R_1$ we have homophily, while at distance scale $\epsilon \sim R_2$ we have heterophily. The horizontal line at $y=0$ corresponds to the point at which the alternative models are more compressive than the purely random mixing model.

Code implementing our method can be found at \texttt{https://github.com/JER-ry/amenities-hom-het}.


\section{Results}
\label{sec:results}

\begin{figure*}
	\centering
	\includegraphics[width=\linewidth]{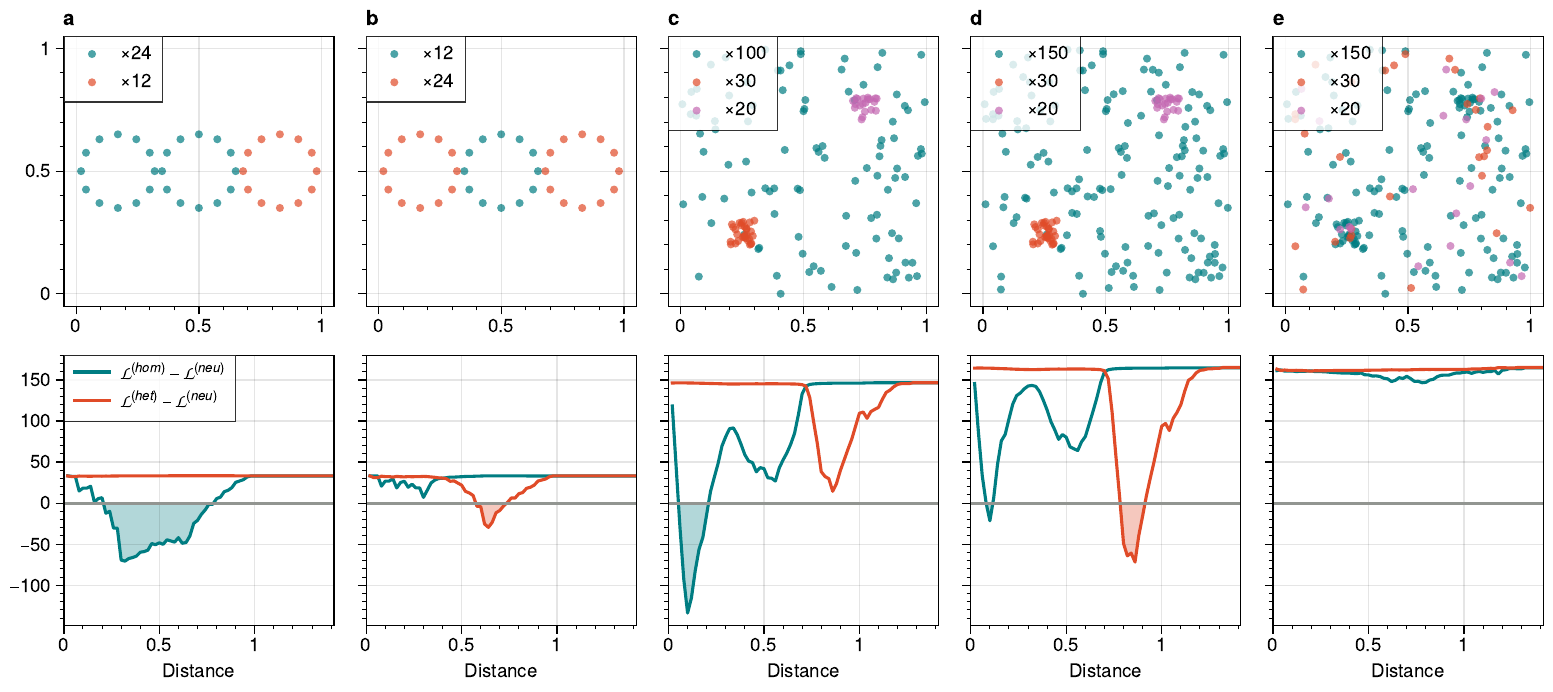}
	\caption{
		\textbf{Synthetic datasets with intuitive mixing structure.} (a) Ring arrangement constructed with strong homophilic mixing at small to moderate distance scales but no meaningful mixing at very small and large scales. (b) Swapping the mesoscale spatial configuration of the rings weakens the homophily at small distance scales, favoring the neutral model, while inducing heterophily at moderate distance scales. (c,d) Uniformly distributed points sampled within small bounding boxes (orange, pink) and the entire domain $[0,1]^2$ (blue), with an increasing frequency of blue points to introduce heterophilic mixing structure at moderate distance scales while weakening the homophilic mixing at small scales. A bimodal regime appears in the homophilic mixing due to the different characteristic scales of clustering. (e) A shuffling of the amenity labels in panel (d) removes all mixing structure so that the neutral model is heavily favored at all distance scales.
	}
	\label{fig:experiment}
\end{figure*}

\subsection{Validation on synthetic data}
\label{sec:synthetic-experiments}

We first validate our framework on synthetic spatial configurations with distinct structural properties. Figure~\ref{fig:experiment} presents two groups of scenarios. The first group (a, b) consists of points arranged in ring structures, while the second group (c--e) features clustered distributions, with (e) serving as a shuffled null model. The node type abundances are presented in the accompanying legends. In scenarios (a) and (b), each of three groups contains $n=12$ points placed at equal angular intervals on a circle of radius $0.15$ centered within the unit square. The three rings are centered at $(0.5, 0.5)$, $(0.17, 0.5)$, and $(0.83, 0.5)$. In (a), the center and left rings share the same label (blue), while the right ring carries a distinct label (orange), creating a segregated arrangement. In (b), the center ring is labeled blue while the left and right rings are both labeled orange, so that the two types alternate across the three circles. In the ring scenarios, the segregated pattern in (a) places points of the same type adjacent to one another along the rings. This concentrated local agglomeration is naturally identified as homophilic at small scales, producing a deep dip in the blue homophily curve, with its minimum occurring at a distance approximately equal to the diameter of one ring. Conversely, the alternating pattern in (b) ensures that nearest neighbors are reliably of different types, which successfully flips the dominant regime to heterophily (orange curve) at scales reflecting this alternating proximity; notably, the minimum of this orange curve occurs at a distance approximately equal to twice the diameter of one ring.

In scenarios (c)--(e), points are drawn from three independent processes using a fixed random seed: 30 points sampled uniformly from $[0.2, 0.3]^2$ (orange), 20 points sampled uniformly from $[0.7, 0.8]^2$ (green), and a set of background points sampled uniformly over $[0,1]^2$ (blue). Scenarios (c) and (d) use 100 and 150 background points, respectively, increasing the relative density of the blue background. Scenario (e) uses the same configuration as (d) but with all type labels randomly permuted, destroying any spatial structure while preserving label frequencies. The highly compact, single-typed clusters in (c) generate an abundance of same-type pairs at very short distances, yielding a strong homophily signal at $\epsilon\sim 0.1$ and a weaker dip at $\epsilon \sim 0.5$ due to the blue background. Scenario (d) maintains these clusters but increases the density of the blue background points, so that while the highly localized homophily from the inner clusters persists, a prominent heterophily signal emerges at intermediate scales. Finally, randomizing the amenity labels in panel (e) eliminates all meaningful spatial structure. As expected, neither model is favored over random mixing, with both curves providing poor compression (e.g. a high description length) across all spatial scales. This confirms that the proposed methodology robustly distinguishes true non-random clustering from spatial noise in a fully nonparametric manner.

\begin{figure*}
	\centering
	\includegraphics[width=\linewidth]{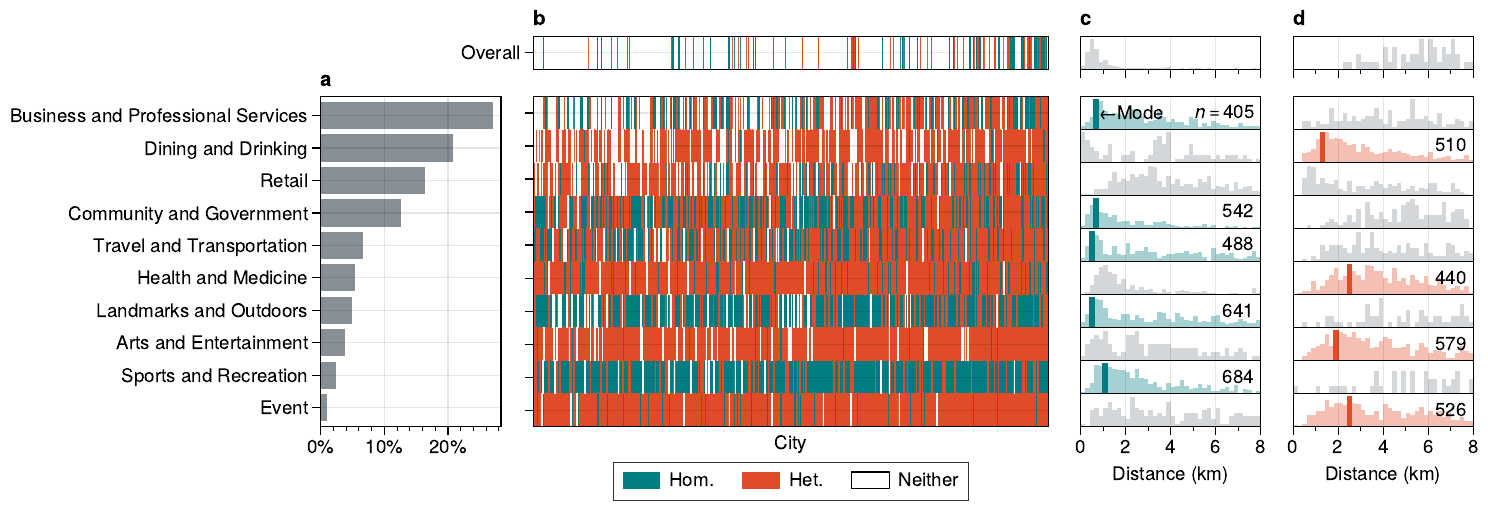}
	\caption{
		\textbf{Universal scales of urban homophily and heterophily across global cities.} (a) Total abundance of each amenity type in the analyzed Foursquare subset \citep{foursquare}, encompassing 38,114,890 amenities in 779 cities worldwide. (b) Mixing classification---homophilic, heterophilic, or neither---extracted by comparing Eqs.~\ref{eq:dl-hom},~\ref{eq:dl-het}~and~\ref{eq:dl-neu} at distance scales from $50$m to $30$km in each city, with every individual city colored according to its mixing classification for each amenity type in panel (a). Amenities exhibit universal mixing tendencies, with homophily or heterophily depending on the amenity type. The mixing tendency across all amenity types, weakened due to the juxtaposition of multiple characteristic clustering scales, is plotted in the bar above. (c,~d) Distributions of the characteristic spatial scales of homophily (c) and heterophily (d) (Eq.~\ref{eq:scale}) associated with each amenity type. Bar counts include only city-category pairs where the respective model (homophily or heterophily) is favored, and bar heights are normalized by the maximum frequency within each type. The modal characteristic scale and the number of cities favoring the model are noted for each amenity type if the model is favored in more than half of the cities. We observe five strongly homophilic amenity types, whose characteristic scales peak at $\epsilon^\ast_{hom}\sim 1$ km, while the other amenity types exhibit strong heterophily that peaks at $\epsilon^\ast_{het}\sim 2$ km (with the exception of `Retail', showing neither mixing tendency). A more fine-grained amenity classification produces even stronger regularities across cities (Fig.~\ref{fig:foursquare_lv2}).    
	}
	\label{fig:results_foursquare}
\end{figure*}

\subsection{Amenity mixing in global cities}
\label{sec:foursquare-results}

To understand the mixing structure of amenities across global cities, we use the Foursquare Open Source Places dataset \citep{foursquare}, a publicly available collection of points of interest (POIs) providing geographic coordinates, venue names, and categorical information for each POI record. The open-source nature of this dataset allows for simple reproduction of our analyses by other researchers using different methodologies and/or urban boundary classifications. Raw POI data are sourced from \citep{wipfliWipfliFoursquareosplacespmtiles2026}, and urban boundaries are drawn from the ESRI World Urban Areas dataset \citep{esriWorldUrbanAreas}. We retain only POIs that carry a category label and fall within an urban boundary, then filter to cities with at least 10,000 POIs, yielding 38,114,890 POIs from 779 cities across 94 countries and territories. POI categories follow the hierarchical taxonomy embedded in the dataset, and for our primary analysis we use the 10 top-level categories: ``Business and Professional Services'', ``Retail'', ``Dining and Drinking'', ``Arts and Entertainment'', ``Community and Government'', ``Health and Medicine'', ``Landmarks and Outdoors'', ``Sports and Recreation'', ``Travel and Transportation'', and ``Event''. The sub-classifications within these top-level amenity classifications are shown in Appendix~\ref{sec:supp-foursquare-data} to provide a more detailed understanding of what's included in each class, while detailed breakdowns of POI counts and category distributions for each city are provided in Table~\ref{tab:foursquare}. Using this dataset, we examine the spatial scales of amenity homophily and heterophily by computing the description lengths of Eqs.~\ref{eq:dl-hom}, \ref{eq:dl-het}, and~\ref{eq:dl-neu} at spatial scales $\epsilon$ ranging from 50~m to 30~km in 50~m increments within each urban boundary. To compute pairwise distances in physically meaningful units while avoiding distortions from global projections, we use local projected coordinates derived from a city-specific map projection for each urban area.

Fig.~\ref{fig:results_foursquare} shows the results of these analyses. In Fig.~\ref{fig:results_foursquare}a we plot the total abundance of each of the 10 top-level amenity types in the dataset. For each city and category, we compare the homophilic, heterophilic, and neutral models using Eqs.~\ref{eq:dl-hom}, \ref{eq:dl-het}, and~\ref{eq:dl-neu} to identify both an overall classification of the amenity type for each city as well as the characteristic scales $\epsilon^\ast$ (Eq.~\ref{eq:scale}) at which these amenities exhibit their most pronounced mixing preferences. A city--category pair is classified as homophilic (heterophilic) if the homophilic (heterophilic) model achieves a lower description length than the neutral model at any scale and also attains a smaller minimum description length over the set of scales examined than the heterophilic (homophilic) model. If a city--category pair is not classified as either homophilic or heterophilic, it is labeled as neither (e.g. neutral), meaning that the purely random mixing model provides a more parsimonious description of the amenity data at all spatial scales studied. To isolate the mixing tendencies of individual amenity types, model selection was performed on each type separately while treating all other types as identical. In other words, when the focus amenity is ``Retail'', all other amenities are given the identical labels ``Not Retail'' to assess the overall extent to which the Retail amenities specifically agglomerate or co-agglomerate. 

In Fig.~\ref{fig:results_foursquare}b we color each individual city according to its mixing classification for the amenity type shown in panel (a). We can see that the amenities exhibit highly shared mixing tendencies, with certain amenities having homophilic preferences across the large majority of cities and others having heterophilic preferences across the large majority of cities. Meanwhile, the mixing tendency when differentiating all amenity types, plotted at the top of panel (b), shows less regularity across cities due to the juxtaposition of multiple characteristic clustering scales. When all the amenity classes are combined---some with strongly homophilic and some with strongly heterophilic mixing preferences---both mixing effects compete at all scales and the neutral model of mixing becomes favorable for many cities. 

Fig.~\ref{fig:results_foursquare}c and Fig.~\ref{fig:results_foursquare}d supplement these classification results with histograms of the characteristic scales (Eq.~\ref{eq:scale}) of homophily and heterophily  respectively, plotted for all cities falling under a given mixing classification. For amenities for which a majority of the cities had the same mixing classification, the corresponding histograms are highlighted with color and the modal distance bin for the characteristic distance scales $\epsilon^\ast$ is marked. We can see that amenities primarily classified as homophilic across cities have a peak scale of mixing at around $\epsilon_{hom}^\ast\sim 1$ km, consistent with neighborhood-level agglomeration \cite{hidalgoAmenityMixUrban2020}, while amenities primarily classified as heterophilic across cities have a peak scale of mixing at around $\epsilon_{het}^\ast\sim 2$ km, reflecting district-level functional complementarity. The scale separation between homophilic and heterophilic interactions is consistent with the rapid spatial decay of agglomeration benefits \citep{arzaghiNetworkingMadisonAvenue2008}, the operation of specialization externalities in very narrow geographies while urbanization externalities extend over larger intra-urban scales \citep{lavoratoriTooCloseComfort2021}, and the zip-code-level co-agglomeration of service industries \citep{kolkoAgglomerationCoAgglomerationServices2007}. The consistency of these characteristic scales across hundreds of cities further suggests that not only the abundance but also the spatial arrangement of amenities exhibits universality across urban contexts, complementing the nonlinear scaling of amenity counts with city population \citep{kaufmannScalingUrbanAmenities2022} by adding a configurational dimension to these regularities. 

Our results identify five homophilic amenity types (``Business and Professional Services'', ``Community and Government'', ``Travel and Transportation'', ``Landmarks and Outdoors'', ``Sports and Recreation'') and four heterophilic amenity types (``Dining and Drinking'', ``Health and Medicine'', ``Arts and Entertainment'', and ``Event''). ``Retail'' is an outlier in these analyses, showing neither mixing tendency, which is perhaps due to the breadth of this category, which encompasses amenities ranging from ``Outlet Mall'' to ``Financial or Legal Service'' (see Appendix~\ref{sec:supp-foursquare-data}).  The category-specific patterns align with the functional logic of each amenity type. Business and professional services, government offices, landmarks, sports venues, and travel hubs---amenities that serve dedicated zones or specialized demand---cluster in a homophilic manner, consistent with zoning-driven concentration. Meanwhile, dining, health, arts, and event amenities, which serve distributed demand and benefit from proximity to complementary services, tend toward heterophily, reflecting co-agglomeration with other amenity types. These patterns are consistent with prior empirical findings that restaurants self-organize into dense spatial clusters when minimum-distance regulations are removed \citep{leonardiAgglomerationUrbanAmenities2023}, and that amenity co-occurrence networks feature neighborhood-scale clusters that align with the peak mixing regimes we detect here \citep{hidalgoAmenityMixUrban2020}. 

As a robustness check, we show that performing label shuffling removes the observed mixing effects (Fig.~\ref{fig:foursquare_shuffle}), which suggests that genuine structural regularities in mixing that move beyond amenity diversity are being identified using the Bayesian methodology. Meanwhile, we show in Fig.~\ref{fig:foursquare_lv2} that a more fine-grained amenity classification produces even stronger regularities across cities. 

\subsection{Longitudinal case study}
\label{sec:hk-restaurants}

The results of the previous section indicate that ``Dining and Drinking''---along with the other leisure-oriented amenities ``Arts \& Entertainment'' and ``Event'' amenities---exhibits heterophilic mixing that peaks at walkable distance scales across hundreds of diverse cities. This consistent spatial pattern likely reflects market-driven benefits for neighborhoods and businesses, prompting a closer, longitudinal examination of how the mixing patterns uncovered by our framework can influence local economic outcomes in the restaurant sector.

\begin{figure*}
	\centering
	\includegraphics[width=\linewidth]{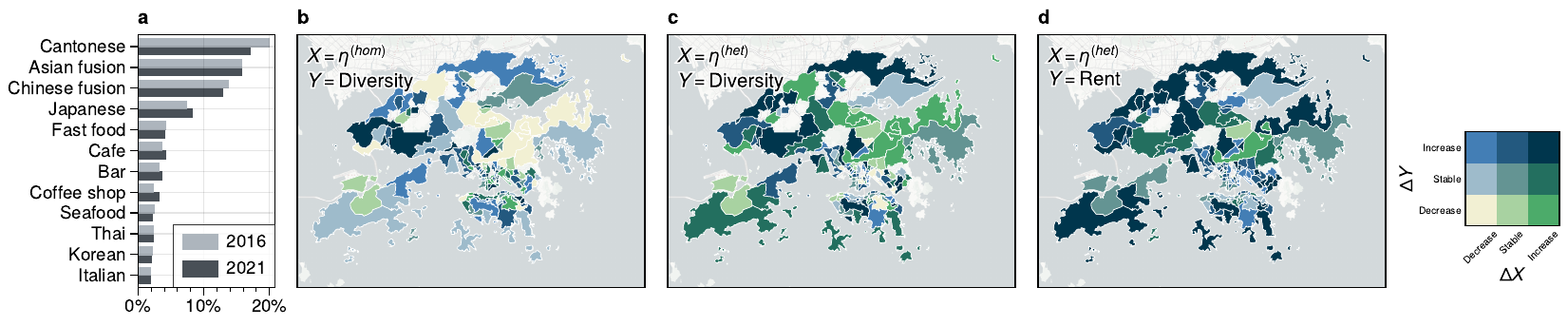}
	\caption{
	\textbf{Longitudinal case study of Hong Kong.}
	(a) Total abundance of cuisine types among active restaurants in 2016 ($N=14003$) and 2021 ($N=16357$), showing only moderate variation across time.
	(b) Bivariate choropleth map plotting the relative change in the homophilic compression ratio (Eq.~\ref{eq:ratio}) at walking scale vs relative change in cuisine diversity (Eq.~\ref{eq:entropy}) from 2016 to 2021 across all 142 studied planning units.
    (c) Relative change in the heterophilic compression ratio at walking scale vs relative change in cuisine diversity.
	(d) Relative change in the heterophilic compression ratio at walking scale versus relative change in median rent from 2016-2021. These plots supplement the regression analysis results in Table~\ref{tab:hk_restaurant_table}. The shared $3\times3$ legend at right encodes the joint $(\Delta X, \Delta Y)$ regime for each LTPUG (see Sec.~\ref{sec:hk-restaurants}).
	}
	\label{fig:results_hk}
\end{figure*}

It has long been argued that urban vitality is linked to density, mixed primary uses, and pedestrian interaction \citep{jacobsDeathLifeGreat1961}, an idea which is supported by substantial empirical work also showing that everyday activity spaces often extend beyond the immediate residential neighborhood \citep{delclsali2019urban,mouratidis2020built}. Service and land-use diversity can be perceived negatively within immediate proximity but positively within walking distance \citep{glaesener2015neighborhood}, and agglomeration effects tend to have greater impact on property markets when building-level specialization matches the surrounding neighborhood \citep{liu2026agglomeration,liu2024agglomeration}. Meanwhile, restaurant clustering can increase demand, reduce search costs, and encourage differentiation  \citep{leonardiAgglomerationUrbanAmenities2023}, while restaurant quality can affect nearby property values \citep{sui2024economic} at the same time it benefits residents beyond the immediate vicinity \citep{kuang2017does,su2022measuring}. Along with data-driven evidence from the analysis of human mobility patterns \citep{juhaszAmenityComplexityUrban2023,heine2025role}, these findings collectively indicate that both the presence and the spatial configuration of diverse amenities matter for local economic development.

These considerations motivate the investigation of how both amenity diversity and spatial mixing impact property values and economic development in cities. The diversity measures commonly used in previous studies capture aggregate compositional information, but neglect information about the spatial configuration of amenities. Hong Kong provides a useful setting for this comparison because its dense development and transit connectivity concentrate activity at walkable scales, making spatial arrangement particularly consequential. Recent evidence shows that walkable transit-oriented areas cluster in central districts and that metro-area vibrancy depends on station accessibility and nearby amenity mix \citep{yu2026characterizing,miao2025vibrancy}. The city's compact form allows us to isolate whether the spatial arrangement of restaurants predicts economic outcomes above and beyond restaurant diversity.

\begin{table*}[!t]
\centering
\caption{OLS estimates of associations among relative changes in cuisine diversity, mixing compression ratios, and median rents across Hong Kong LTPUGs from 2016–2021.
}
\label{tab:hk_restaurant_table}
\begin{threeparttable}
\scriptsize
\setlength{\tabcolsep}{10pt}
\renewcommand{\arraystretch}{1.24}
\setlength{\heavyrulewidth}{0.10em}
\setlength{\lightrulewidth}{0.06em}
\setlength{\cmidrulewidth}{0.06em}

\begin{tabular}{c l c c c c c}
\toprule
\multirow{2}{*}{Specification}
& \multirow{2}{*}{Covariates}
& \multicolumn{2}{c}{Unstandardized Coef.}
& \multicolumn{1}{c}{Standardized Coef.}
& \multirow{2}{*}{$t$-statistic}
& \multirow{2}{*}{$p$-value} \\
\cmidrule(r{1.0em}){3-4}\cmidrule(l{1.0em}){5-5}
&  & $B$ & Std. Err. & $\beta$ &  &  \\
\midrule

\multirow{2}{*}{\makecell[c]{$\Delta \mathrm{Rent}^{2021\text{--}2016}$\\$\sim \Delta H^{(\mathrm{cuisine}),\,2021\text{--}2016}$}}
& (Constant) & 26.9758 & 1.9490 &  & 13.839 &  \\
& $\%$ Change in Diversity & 0.6849 & 0.5067 & 0.1135 & 1.352 & 0.179 \\
\midrule

\multirow{2}{*}{\makecell[c]{$\Delta \mathrm{Rent}^{2021\text{--}2016}$\\$\sim \Delta \eta^{(hom),\,2021\text{--}2016}$}}
& (Constant) & 27.3752 & 1.8920 &  & 14.467 &  \\
& $\%$ Change in Homophily & $-105.5638^{*}$ & 43.0305 & $-0.2030^{*}$ & $-2.453$ & 0.015 \\
\midrule

\multirow{2}{*}{\makecell[c]{$\Delta \mathrm{Rent}^{2021\text{--}2016}$\\$\sim \Delta \eta^{(het),\,2021\text{--}2016}$}}
& (Constant) & 24.4582 & 2.0120 &  & 12.154 &  \\
& $\%$ Change in Heterophily & $1291.9229^{***}$ & 349.5870 & $0.2981^{***}$ & 3.696 & $<0.001$ \\
\midrule

\multirow{3}{*}{\makecell[c]{$\Delta \mathrm{Rent}^{2021\text{--}2016}$\\$\sim \Delta H^{(\mathrm{cuisine}),\,2021\text{--}2016}$\\$+\ \Delta \eta^{(hom),\,2021\text{--}2016}$}}
& (Constant) & 26.8470 & 1.9120 &  & 14.045 &  \\
& $\%$ Change in Diversity & 0.7880 & 0.4984 & 0.1306 & 1.581 & 0.116 \\
& $\%$ Change in Homophily & $-111.0003^{*}$ & 42.9397 & $-0.2135^{*}$ & $-2.585$ & 0.011 \\
\midrule

\multirow{3}{*}{\makecell[c]{$\Delta \mathrm{Rent}^{2021\text{--}2016}$\\$\sim \Delta H^{(\mathrm{cuisine}),\,2021\text{--}2016}$\\$+\ \Delta \eta^{(het),\,2021\text{--}2016}$}}
& (Constant) & 23.6018 & 2.0390 &  & 11.572 &  \\
& $\%$ Change in Diversity & $0.9546$ & 0.4867 & $0.1582$ & 1.961 & 0.052 \\
& $\%$ Change in Heterophily & $1387.5770^{***}$ & 349.5068 & $0.3202^{***}$ & 3.970 & $<0.001$ \\

\bottomrule
\end{tabular}

\begin{tablenotes}[flushleft]
\footnotesize
\item All models use $N=142$ TPUs. Reported $p$-values are from two-sided $t$-tests with 140 degrees of freedom in the univariate models and 139 degrees of freedom in the bivariate models. $^{***}$, $^{**}$, and $^{*}$ indicate statistical significance at the 0.1\%, 1\%, and 5\% levels, respectively.
\end{tablenotes}

\end{threeparttable}
\end{table*}

We explore the relationships among restaurant diversity, spatial mixing, and local property values through a longitudinal case study of restaurants across Hong Kong. We construct a monthly panel dataset using restaurant licenses obtained from the Hong Kong Food and Environmental Hygiene Department (FEHD) from January 2016 through December 2021 \citep{hkfehdRestaurantLicences2026}, assigning each restaurant spatial coordinates by merging FEHD processed geospatial license files where available and geocoding unmatched addresses using Google Maps \citep{googleGeocodingAPI2026}. Cuisine labels are assigned by first matching restaurants to Google Places using name and address queries and mapping the returned place types to a standardized cuisine classification \citep{googlePlaceAPI2026}. Records that remain unresolved, or that are tagged only with generic types, are supplemented by name- and address-based classification using Gemini 2.5 Flash \citep{gemini2025}. We spatially join restaurants to Large Tertiary Planning Unit Groups (LTPUGs) and attach LTPUG-level census attributes, including median monthly domestic household rent, from the 2016 and 2021 Population Census \citep{hkcensusLTPUG2016,hkcensusLTPUG2021}. The dataset includes 142 LTPUGs that are consistent across 2016 and 2021. We average monthly LTPUG cuisine profiles over June to August 2016 and June to August 2021 to align the restaurant measures with the census reference period. This process gives us two spatially resolved restaurant datasets $\{\bm{X},\bm{b}\}$, one for 2016 ($N=14003$ restaurants) and one for 2021 ($N=16357$ restaurants), with each restaurant categorized by its cuisine type.

Cuisine diversity within each LTPUG is measured by normalized Shannon entropy,
\begin{align}
H_{t}^{(\mathrm{cuisine})} = -\frac{1}{\log K_t}\sum_{c=1}^{K_t} p_{c,t} \log p_{c,t},
\label{eq:entropy}
\end{align}
where $p_{c,t}$ is the restaurant share of cuisine $c$ in year $t$, $K_t > 1$ is the number of cuisines with positive counts in that LTPUG in year $t$, and $t \in \{2016,2021\}$. We compare this diversity measure with our proposed framework, which is applied at a neighborhood scale of $\epsilon=900\mathrm{m}$ within each LTPUG---approximately a 10-minute walk and the point at which the homophily and heterophily distributions across LTPUGs stabilize (see Fig.~\ref{fig:cr_across_scales}).  Figure~\ref{fig:results_hk} provides a visual comparison of both methods alongside the extracted median rent values for each LTPUG using bivariate choropleth maps.

Figure~\ref{fig:results_hk}a shows the market shares of the most common cuisine types in our dataset, indicating the slow evolution of restaurant composition over time. Figures~\ref{fig:results_hk}(b--d) show bivariate choropleth maps of the diversity, compression ratios, and median rents across the city in different combinations. On each percentage-change axis $\Delta X$ and $\Delta Y$, LTPUGs are partitioned into one of three regimes (Decrease, Stable, Increase) for the visualization, using a tertile scheme centered at zero. Percentage changes below zero are split at their $67$th percentile and the above-zero values are split at their $33$rd percentile, so that the ``Decrease'' regime covers the bottom two thirds of negative values, ``Increase'' covers the top two thirds of positive values, and ``Stable'' is the middle band combining the top third of negative values with the bottom third of positive values. These maps show that changes in cuisine diversity alone do not uniquely determine the spatial form of neighborhood change: rising diversity can coexist with stable-to-increasing homophily, but it aligns more clearly with stable-to-increasing heterophily. The contrast indicates that similar compositional diversification can reflect either homophilic or heterophilic mixing. Panel (d) links these structural signals to economic outcomes, showing that high rent growth LTPUGs are disproportionately concentrated among those with rising heterophily.

To more rigorously determine these associations, we regress the relative change in median rent on relative changes in cuisine diversity and both compression ratios, in different combinations. Table~\ref{tab:hk_restaurant_table} reports the OLS results across different model specifications. Increases in the heterophily compression ratio are a robust positive predictor of rent growth, while homophily has a weak negative association with rent changes in different settings. Cuisine diversity, on the other hand, is not a statistically significant predictor of rent growth, either in isolation or in conjunction with the spatial mixing measures. These results extend prior work on the capitalization of local retail and service environments by showing that the spatial arrangement of amenities matters alongside their composition \citep{ghorbaniAreLocalRetail2024,glaeserNowcastingGentrificationUsing2018}, and suggest the importance of heterophilic mixing at walkable scales for local urban vitality.


\section{Conclusion}

Understanding the agglomerative and co-agglomerative forces that underlie the spatial distribution of urban amenities is important for building a general science of cities. By treating spatial mixing as a model selection problem grounded in the MDL principle, here we enable a fair cross-sectional analysis of amenity agglomeration and co-agglomeration tendencies across almost 800 cities worldwide, in order to understand the extent to which universal mechanisms shape these urban structural patterns. Casting agglomeration as Bayesian model selection over proximity network generative processes provides a principled answer for whether an observed spatial arrangement constitutes statistical evidence for attraction or repulsion between different amenities, and to what extent such a structured model of agglomeration---homophilic (attractive) or heterophilic (complementary)---is favored over a neutral (unstructured) model of spatial mixing. The direct mapping between this network generative model and a utility maximizing location choice process (see Appendix~\ref{sec:physical}) further grounds our statistical framework in economic theory, connecting information theoretic parsimony to the microeconomic mechanisms that underlie agglomeration.

The empirical results we find by applying this framework across 779 cities reveal a high level of regularity in spatial mixing among different amenity types. The mixing regime of an amenity category (homophilic or heterophilic) depends largely on the amenity's functional role rather than city size, geography, or cultural context. Business, government, and travel-related establishments cluster homophilically at walkable distances, consistent with coordination and information sharing benefits that require physical co-presence. Dining, health, and entertainment amenities exhibit heterophily at comparable scales, consistent with consumer demand for variety and complementary services within a single trip. These patterns emerge at similar spatial scales across cities with vastly different institutional and urban morphological contexts. This consistency supports the view that certain spatial regularities of economic organization are genuinely universal, more akin to scaling laws of urbanization than locally contingent planning outcomes. Meanwhile, the prominent heterophilic mixing of certain amenities across hundreds of diverse cities points to market-driven neighborhood and business benefits of co-agglomeration of specific amenities at walkable scales. To more closely examine how this small-scale heterophilic mixing relates to neighborhood-level economic outcomes, our longitudinal Hong Kong case study follows up to find that changes in heterophilic mixing among restaurants predict subsequent neighborhood rent growth, whereas compositional diversity of cuisine types does not. The structural regularities inferred through the proposed framework therefore encode important information about economic vitality that aggregate diversity measures systematically discard.

There are several directions for future work that could strengthen our empirical findings. Incorporating temporal dynamics more systematically would allow tracking of how mixing regimes shift in response to zoning reforms, infrastructure investment, or economic shocks. Providing a causal identification of the mechanisms driving agglomeration is required for solidifying a theory of spatial mixing for a unified urban science. It is also important to examine the mixing structure present at sub-city scales, as done in the Hong Kong case study, to permit detection of intra-city variation in such regularities and its relationship to land use policy. 

There are also a number of ways to further apply the proposed Bayesian methodology to identify mixing structure relevant to other urban processes. For instance, the framework could be used to analyze spatial mixing of native and invasive species across urban green spaces, revealing whether biodiversity patterns follow functional rules similar to amenity agglomeration. It could also be adapted to measure residential segregation by race or income at fine spatial scales, providing a parameter-free alternative to traditional dissimilarity indices that avoids arbitrary neighborhood boundaries. Similarly, the method could evaluate land use mixing patterns, such as the spatial arrangement of residential, commercial, and industrial zones, and test whether observed mixing regimes align with planning objectives or emergent market dynamics.


\section*{Acknowledgments}
\vspace{-\baselineskip}
This work was supported in part by the Hong Kong Research Grants Council through General Research Fund Project No. 17301024 and through the HKU-100 Start Up Fund (AK). The authors thank Shihui Feng for helpful discussions and Mohsen Rahimi for collecting a preliminary version of the Hong Kong restaurant dataset.

\bibliographystyle{numeric}
\bibliography{refs}

@article{ahlfeldtEconomicsDensityEvidence2015,
  title = {The Economics of Density: Evidence From the {{Berlin Wall}}},
  author = {Ahlfeldt, Gabriel M. and Redding, Stephen J. and Sturm, Daniel M. and Wolf, Nikolaus},
  year = 2015,
  month = nov,
  journal = {Econometrica},
  volume = {83},
  number = {6},
  pages = {2127--2189},
  publisher = {John Wiley \& Sons, Ltd},
  issn = {0012-9682},
  doi = {10.3982/ECTA10876},
  urldate = {2026-03-11}
}

@article{kirkley2020information,
  title={Information theoretic network approach to socioeconomic correlations},
  author={Kirkley, Alec},
  journal={Physical Review Research},
  volume={2},
  number={4},
  pages={043212},
  year={2020},
  publisher={APS}
}

@article{youn2016scaling,
  title={Scaling and universality in urban economic diversification},
  author={Youn, Hyejin and Bettencourt, Lu{\'\i}s and Lobo, Jos{\'e} and Strumsky, Deborah and Samaniego, Horacio and West, Geoffrey B},
  journal={Journal of The Royal Society Interface},
  volume={13},
  number={114},
  year={2016},
  publisher={The Royal Society}
}

@article{kirkley2025fast,
  title = {Fast nonparametric inference of network backbones for weighted graph sparsification},
  author = {Kirkley, Alec},
  journal = {Physical Review X},
  volume = {15},
  number = {3},
  pages = {031013},
  year = {2025},
  publisher = {APS}
}

@inproceedings{kontkanen2007mdl,
  title = {{{MDL}} histogram density estimation},
  author = {Kontkanen, Petri and Myllym{\"a}ki, Petri},
  booktitle = {Artificial intelligence and statistics},
  pages = {219--226},
  year = {2007},
  organization = {PMLR}
}

@article{yang2023unsupervised,
  title = {Unsupervised discretization by two-dimensional {{MDL}}-based histogram},
  author = {Yang, Lincen and Baratchi, Mitra and van Leeuwen, Matthijs},
  journal = {Machine Learning},
  volume = {112},
  number = {7},
  pages = {2397--2431},
  year = {2023},
  publisher = {Springer}
}

@inproceedings{papadimitriou2005parameter,
  title = {Parameter-free spatial data mining using {{MDL}}},
  author = {Papadimitriou, Spiros and Gionis, Aristides and Tsaparas, Panayiotis and Vaisanen, Risto A and Mannila, Heikki and Faloutsos, Christos},
  booktitle = {Fifth IEEE International Conference on Data Mining (ICDM'05)},
  pages = {8--pp},
  year = {2005},
  organization = {IEEE}
}

@inproceedings{coupette2022differentially,
  title = {Differentially describing groups of graphs},
  author = {Coupette, Corinna and Dalleiger, Sebastian and Vreeken, Jilles},
  booktitle = {Proceedings of the AAAI Conference on Artificial Intelligence},
  volume = {36},
  number = {4},
  pages = {3959--3967},
  year = {2022}
}

@article{peixoto2022ordered,
  title = {Ordered community detection in directed networks},
  author = {Peixoto, Tiago P},
  journal = {Physical Review E},
  volume = {106},
  number = {2},
  pages = {024305},
  year = {2022},
  publisher = {APS}
}

@article{kirkley2025structural,
  title = {Structural reducibility of hypergraphs},
  author = {Kirkley, Alec and Felippe, Helcio and Battiston, Federico},
  journal = {Physical Review Letters},
  volume = {135},
  number = {24},
  pages = {247401},
  year = {2025},
  publisher = {APS}
}

@article{morel2024bayesian,
  title = {{{Bayesian}} regionalization of urban mobility networks},
  author = {Morel-Balbi, Sebastian and Kirkley, Alec},
  journal = {Physical Review Research},
  volume = {6},
  number = {3},
  pages = {033307},
  year = {2024},
  publisher = {APS}
}

@article{kirkley2022spatial,
  title = {Spatial regionalization based on optimal information compression},
  author = {Kirkley, Alec},
  journal = {Communications Physics},
  volume = {5},
  number = {1},
  pages = {249},
  year = {2022},
  publisher = {Nature Publishing Group UK London}
}

@article{kirkley2024identifying,
  title = {Identifying hubs in directed networks},
  author = {Kirkley, Alec},
  journal = {Physical Review E},
  volume = {109},
  number = {3},
  pages = {034310},
  year = {2024},
  publisher = {APS}
}

@article{gallagher2021clarified,
  title = {A clarified typology of core-periphery structure in networks},
  author = {Gallagher, Ryan J and Young, Jean-Gabriel and Welles, Brooke Foucault},
  journal = {Science advances},
  volume = {7},
  number = {12},
  pages = {eabc9800},
  year = {2021},
  publisher = {American Association for the Advancement of Science}
}

@article{ripley1977modelling,
  title = {Modelling spatial patterns},
  author = {Ripley, Brian D},
  journal = {Journal of the Royal Statistical Society: Series B (Methodological)},
  volume = {39},
  number = {2},
  pages = {172--192},
  year = {1977},
  publisher = {Wiley Online Library}
}

@article{barthelemy2011spatial,
  title = {Spatial networks},
  author = {Barth{\'e}lemy, Marc},
  journal = {Physics Reports},
  volume = {499},
  number = {1-3},
  pages = {1--101},
  year = {2011},
  publisher = {Elsevier}
}

@article{leslie2011colocation,
  title = {The Colocation Quotient: A New Measure of Spatial Association Between Categorical Subsets of Points},
  author = {Leslie, Timothy F and Kronenfeld, Barry J},
  journal = {Geographical Analysis},
  volume = {43},
  number = {3},
  pages = {306--326},
  year = {2011}
}

@article{cromley2014geographically,
  title = {Geographically weighted colocation quotients: specification and application},
  author = {Cromley, Robert G and Hanink, Dean M and Bentley, George C},
  journal = {The Professional Geographer},
  volume = {66},
  number = {1},
  pages = {138--148},
  year = {2014},
  publisher = {Taylor \& Francis}
}

@article{marcon2013statistical,
  title = {A statistical test for {Ripley}'s K function rejection of {Poisson} null hypothesis},
  author = {Marcon, Eric and Traissac, St{\'e}phane and Lang, Gabriel},
  journal = {International Scholarly Research Notices},
  volume = {2013},
  number = {1},
  pages = {753475},
  year = {2013},
  publisher = {Wiley Online Library}
}

@article{peixoto2019bayesian,
  title = {{{Bayesian}} stochastic blockmodeling},
  author = {Peixoto, Tiago P},
  journal = {Advances in network clustering and blockmodeling},
  pages = {289--332},
  year = {2019},
  publisher = {Wiley Online Library}
}

@article{peixoto2017nonparametric,
  title = {Nonparametric {{Bayesian}} inference of the microcanonical stochastic block model},
  author = {Peixoto, Tiago P},
  journal = {Physical Review E},
  volume = {95},
  number = {1},
  pages = {012317},
  year = {2017},
  publisher = {APS}
}

@article{morel2025estimation,
  title = {Estimation of partial rankings from sparse, noisy comparisons},
  author = {Morel-Balbi, Sebastian and Kirkley, Alec},
  journal = {Communications Physics},
  volume = {9},
  pages = {30},
  year = {2026},
  publisher = {Nature Publishing Group UK London}
}

@article{rissanen1978modeling,
  title = {Modeling by shortest data description},
  author = {Rissanen, Jorma},
  journal = {Automatica},
  volume = {14},
  number = {5},
  pages = {465--471},
  year = {1978},
  publisher = {Elsevier}
}

@article{kirkley2023compressing,
  title = {Compressing network populations with modal networks reveal structural diversity},
  author = {Kirkley, Alec and Rojas, Alexis and Rosvall, Martin and Young, Jean-Gabriel},
  journal = {Communications Physics},
  volume = {6},
  number = {1},
  pages = {148},
  year = {2023},
  publisher = {Nature Publishing Group UK London}
}

@article{arcauteConstructingCitiesDeconstructing2015,
  title = {Constructing Cities, Deconstructing Scaling Laws},
  author = {Arcaute, Elsa and Hatna, Erez and Ferguson, Peter and Youn, Hyejin and Johansson, Anders and Batty, Michael},
  year = 2015,
  month = jan,
  journal = {Journal of The Royal Society Interface},
  volume = {12},
  number = {102},
  pages = {20140745},
  issn = {1742-5689},
  doi = {10.1098/rsif.2014.0745},
  urldate = {2026-03-13}
}

@article{arzaghiNetworkingMadisonAvenue2008,
  title = {Networking off {{Madison Avenue}}},
  author = {Arzaghi, Mohammad and Henderson, J. Vernon},
  year = 2008,
  month = oct,
  journal = {The Review of Economic Studies},
  volume = {75},
  number = {4},
  pages = {1011--1038},
  issn = {0034-6527},
  doi = {10.1111/j.1467-937X.2008.00499.x},
  urldate = {2026-03-10}
}

@article{battyBuildingScienceCities2012,
  title = {Building a Science of Cities},
  author = {Batty, Michael},
  year = 2012,
  month = mar,
  journal = {Cities},
  series = {Current {{Research}} on {{Cities}}},
  volume = {29},
  pages = {S9-S16},
  issn = {0264-2751},
  doi = {10.1016/j.cities.2011.11.008},
  urldate = {2026-03-24}
}

@book{battyFractalCitiesGeometry1994,
  title = {Fractal Cities: A Geometry of Form and Function},
  shorttitle = {Fractal Cities},
  author = {Batty, Michael and Longley, Paul},
  year = 1994,
  month = jul,
  publisher = {Academic Press Professional, Inc.},
  address = {USA},
  isbn = {978-0-12-455570-9}
}

@book{battyNewScienceCities2013,
  title = {The {{New Science}} of {{Cities}}},
  author = {Batty, Michael},
  year = 2013,
  month = nov,
  publisher = {The MIT Press},
  doi = {10.7551/mitpress/9399.001.0001},
  urldate = {2026-03-14},
  isbn = {978-0-262-31823-5}
}

@article{bettencourtGrowthInnovationScaling2007,
  title = {Growth, Innovation, Scaling, and the Pace of Life in Cities},
  author = {Bettencourt, Lu{\'i}s M. A. and Lobo, Jos{\'e} and Helbing, Dirk and K{\"u}hnert, Christian and West, Geoffrey B.},
  year = 2007,
  month = apr,
  journal = {Proceedings of the National Academy of Sciences},
  volume = {104},
  number = {17},
  pages = {7301--7306},
  publisher = {Proceedings of the National Academy of Sciences},
  doi = {10.1073/pnas.0610172104},
  urldate = {2026-03-14}
}

@article{bettencourtOriginsScalingCities2013,
  title = {The Origins of Scaling in Cities},
  author = {Bettencourt, Lu{\'i}s M. A.},
  year = 2013,
  month = jun,
  journal = {Science},
  volume = {340},
  number = {6139},
  pages = {1438--1441},
  publisher = {American Association for the Advancement of Science},
  doi = {10.1126/science.1235823},
  urldate = {2026-03-14}
}

@article{billingsAgglomerationUrbanArea2016,
  title = {Agglomeration within an Urban Area},
  author = {Billings, Stephen B. and Johnson, Erik B.},
  year = 2016,
  month = jan,
  journal = {Journal of Urban Economics},
  volume = {91},
  pages = {13--25},
  issn = {0094-1190},
  doi = {10.1016/j.jue.2015.11.002},
  urldate = {2026-03-14}
}

@article{bryanCitiesDevelopingWorld2020,
  title = {Cities in the Developing World},
  author = {Bryan, Gharad and Glaeser, Edward and Tsivanidis, Nick},
  year = 2020,
  month = aug,
  journal = {Annual Review of Economics},
  volume = {12},
  number = {Volume 12, 2020},
  pages = {273--297},
  publisher = {Annual Reviews},
  issn = {1941-1383, 1941-1391},
  doi = {10.1146/annurev-economics-080218-030303},
  urldate = {2026-03-12}
}

@incollection{carlinoChapter6Agglomeration2015,
  title = {Chapter 6 - {{Agglomeration}} and {{Innovation}}},
  booktitle = {Handbook of {{Regional}} and {{Urban Economics}}},
  author = {Carlino, Gerald and Kerr, William R.},
  editor = {Duranton, Gilles and Henderson, J. Vernon and Strange, William C.},
  year = 2015,
  month = jan,
  volume = {5},
  pages = {349--404},
  publisher = {Elsevier},
  doi = {10.1016/B978-0-444-59517-1.00006-4},
  isbn = {1574-0080}
}

@article{chinIdentifyingUrbanFunctional2024,
  title = {Identifying Urban Functional Zones by Analysing the Spatial Distribution of Amenities},
  author = {Chin, Wei Chien Benny and Fu, Yuming and Lim, Kwan Hui and Schroepfer, Thomas and Cheah, Lynette},
  year = 2024,
  month = jul,
  journal = {Environment and Planning B: Urban Analytics and City Science},
  volume = {51},
  number = {6},
  pages = {1274--1289},
  issn = {2399-8083, 2399-8091},
  doi = {10.1177/23998083231217376},
  urldate = {2026-01-11}
}

@incollection{combesChapter5Empirics2015,
  title = {Chapter 5 - {{The Empirics}} of {{Agglomeration Economies}}},
  booktitle = {Handbook of {{Regional}} and {{Urban Economics}}},
  author = {Combes, Pierre-Philippe and Gobillon, Laurent},
  editor = {Duranton, Gilles and Henderson, J. Vernon and Strange, William C.},
  year = 2015,
  month = jan,
  volume = {5},
  pages = {247--348},
  publisher = {Elsevier},
  doi = {10.1016/B978-0-444-59517-1.00005-2},
  isbn = {1574-0080}
}

@incollection{durantonChapter48MicroFoundations2004,
  title = {Chapter 48 - {{Micro-Foundations}} of {{Urban Agglomeration Economies}}},
  booktitle = {Handbook of {{Regional}} and {{Urban Economics}}},
  author = {Duranton, Gilles and Puga, Diego},
  editor = {Henderson, J. Vernon and Thisse, Jacques-Fran{\c c}ois},
  year = 2004,
  month = jan,
  volume = {4},
  pages = {2063--2117},
  publisher = {Elsevier},
  doi = {10.1016/S1574-0080(04)80005-1},
  isbn = {1574-0080}
}

@article{ellisonWhatCausesIndustry2010,
  title = {What causes industry agglomeration? {{Evidence}} from coagglomeration patterns},
  shorttitle = {What {{Causes Industry Agglomeration}}?},
  author = {Ellison, Glenn and Glaeser, Edward L. and Kerr, William R.},
  year = 2010,
  month = jun,
  journal = {American Economic Review},
  volume = {100},
  number = {3},
  pages = {1195--1213},
  issn = {0002-8282},
  doi = {10.1257/aer.100.3.1195},
  urldate = {2026-03-14}
}

@article{fotheringhamModifiableArealUnit1991,
  title = {The modifiable areal unit problem in multivariate statistical analysis},
  author = {Fotheringham, A S and Wong, D W S},
  year = 1991,
  month = jul,
  journal = {Environment and Planning A: Economy and Space},
  volume = {23},
  number = {7},
  pages = {1025--1044},
  publisher = {SAGE Publications Ltd},
  issn = {0308-518X},
  doi = {10.1068/a231025},
  urldate = {2026-03-12}
}

@article{ghorbaniAreLocalRetail2024,
  title = {Are Local Retail Services an Amenity or a Nuisance?},
  author = {Ghorbani, Pooya and Meltzer, Rachel},
  year = 2024,
  month = mar,
  journal = {Journal of Economic Geography},
  volume = {24},
  number = {2},
  pages = {193--218},
  issn = {1468-2702, 1468-2710},
  doi = {10.1093/jeg/lbad034},
  urldate = {2025-12-17}
}

@article{giulianoAgglomerationEconomiesEvolving2019,
  title = {Agglomeration Economies and Evolving Urban Form},
  author = {Giuliano, Genevieve and Kang, Sanggyun and Yuan, Quan},
  year = 2019,
  month = dec,
  journal = {The Annals of Regional Science},
  volume = {63},
  number = {3},
  pages = {377--398},
  issn = {1432-0592},
  doi = {10.1007/s00168-019-00957-4}
}

@book{glaeserCitiesAgglomerationSpatial2008,
  title = {Cities, {{Agglomeration}}, and {{Spatial Equilibrium}}},
  author = {Glaeser, Edward L.},
  year = 2008,
  month = jul,
  series = {The {{Lindahl Lectures}}},
  publisher = {Oxford University Press},
  address = {Oxford, New York},
  isbn = {978-0-19-929044-4}
}

@incollection{glaeserIntroduction2010,
  title = {Introduction},
  booktitle = {Agglomeration {{Economics}}},
  author = {Glaeser, Edward L.},
  year = 2010,
  month = feb,
  pages = {1--14},
  publisher = {University of Chicago Press},
  urldate = {2026-01-11}
}

@article{glaeserNowcastingGentrificationUsing2018,
  title = {Nowcasting gentrification: using {{Yelp}} data to quantify neighborhood change},
  shorttitle = {Nowcasting {{Gentrification}}},
  author = {Glaeser, Edward L. and Kim, Hyunjin and Luca, Michael},
  year = 2018,
  month = may,
  journal = {AEA Papers and Proceedings},
  volume = {108},
  pages = {77--82},
  issn = {2574-0768},
  doi = {10.1257/pandp.20181034},
  urldate = {2026-03-14}
}

@article{gonzalezUnderstandingIndividualHuman2008,
  title = {Understanding Individual Human Mobility Patterns},
  author = {Gonz{\'a}lez, Marta C. and Hidalgo, C{\'e}sar A. and Barab{\'a}si, Albert-L{\'a}szl{\'o}},
  year = 2008,
  month = jun,
  journal = {Nature},
  volume = {453},
  number = {7196},
  pages = {779--782},
  publisher = {Nature Publishing Group},
  issn = {1476-4687},
  doi = {10.1038/nature06958},
  urldate = {2026-03-14}
}

@article{hidalgoAmenityMixUrban2020,
  title = {The Amenity Mix of Urban Neighborhoods},
  author = {Hidalgo, C{\'e}sar A. and Casta{\~n}er, Elisa and Sevtsuk, Andres},
  year = 2020,
  month = dec,
  journal = {Habitat International},
  volume = {106},
  pages = {102205},
  issn = {01973975},
  doi = {10.1016/j.habitatint.2020.102205},
  urldate = {2025-12-17}
}

@article{juhaszAmenityComplexityUrban2023,
  title = {Amenity Complexity and Urban Locations of Socio-Economic Mixing},
  author = {Juh{\'a}sz, S{\'a}ndor and Pint{\'e}r, Gerg{\H o} and Kov{\'a}cs, {\'A}d{\'a}m J. and Borza, Endre and M{\'o}nus, Gergely and L{\H o}rincz, L{\'a}szl{\'o} and Lengyel, Bal{\'a}zs},
  year = 2023,
  month = sep,
  journal = {EPJ Data Science},
  volume = {12},
  number = {1},
  pages = {34},
  issn = {2193-1127},
  doi = {10.1140/epjds/s13688-023-00413-6},
  urldate = {2026-01-11}
}

@article{kerrAgglomerativeForcesCluster2015,
  title = {Agglomerative forces and cluster shapes},
  author = {Kerr, William R. and Kominers, Scott Duke},
  year = 2015,
  month = oct,
  journal = {The Review of Economics and Statistics},
  volume = {97},
  number = {4},
  pages = {877--899},
  issn = {0034-6535},
  doi = {10.1162/REST_a_00471},
  urldate = {2026-03-11}
}

@misc{kolkoAgglomerationCoAgglomerationServices2007,
  type = {{SSRN Scholarly Paper}},
  title = {Agglomeration and co-agglomeration of services industries},
  author = {Kolko, Jed},
  year = 2007,
  month = apr,
  number = {985711},
  eprint = {985711},
  publisher = {Social Science Research Network},
  address = {Rochester, NY},
  doi = {10.2139/ssrn.985711},
  urldate = {2026-03-14},
  archiveprefix = {Social Science Research Network}
}

@article{krugmanIncreasingReturnsEconomic1991,
  title = {Increasing returns and economic geography},
  author = {Krugman, Paul},
  year = 1991,
  journal = {Journal of Political Economy},
  volume = {99},
  number = {3},
  eprint = {2937739},
  eprinttype = {jstor},
  pages = {483--499},
  publisher = {University of Chicago Press},
  issn = {00223808, 1537534X},
  urldate = {2026-03-13}
}

@article{lavoratoriTooCloseComfort2021,
  title = {Too close for comfort? {{Microgeography}} of agglomeration economies in the {United Kingdom}},
  shorttitle = {Too Close for Comfort?},
  author = {Lavoratori, Katiuscia and Castellani, Davide},
  year = 2021,
  journal = {Journal of Regional Science},
  volume = {61},
  number = {5},
  pages = {1002--1028},
  issn = {1467-9787},
  doi = {10.1111/jors.12531},
  urldate = {2026-03-12}
}

@article{leonardiAgglomerationUrbanAmenities2023,
  title = {The agglomeration of urban amenities: evidence from {{Milan}} restaurants},
  shorttitle = {The {{Agglomeration}} of {{Urban Amenities}}},
  author = {Leonardi, Marco and Moretti, Enrico},
  year = 2023,
  month = jun,
  journal = {American Economic Review: Insights},
  volume = {5},
  number = {2},
  pages = {141--157},
  issn = {2640-205X, 2640-2068},
  doi = {10.1257/aeri.20220011},
  urldate = {2025-12-17}
}

@book{marshallPrinciplesEconomics1890,
  title = {Principles of {{Economics}}},
  author = {Marshall, Alfred},
  year = 1890,
  publisher = {Macmillan}
}

@incollection{mccannTheoriesAgglomerationRegional2019,
  title = {Theories of Agglomeration and Regional Economic Growth: A Historical Review},
  shorttitle = {Theories of Agglomeration and Regional Economic Growth},
  booktitle = {Handbook of {{Regional Growth}} and {{Development Theories}}},
  author = {McCann, Philip and van Oort, Frank},
  year = 2019,
  month = jul,
  pages = {6--23},
  publisher = {Edward Elgar Publishing},
  urldate = {2025-12-17},
  chapter = {Handbook of Regional Growth and Development Theories},
  isbn = {978-1-78897-002-0}
}

@incollection{neumarkChapter18PlaceBased2015,
  title = {Chapter 18 - {{Place-Based Policies}}},
  booktitle = {Handbook of {{Regional}} and {{Urban Economics}}},
  author = {Neumark, David and Simpson, Helen},
  editor = {Duranton, Gilles and Henderson, J. Vernon and Strange, William C.},
  year = 2015,
  month = jan,
  volume = {5},
  pages = {1197--1287},
  publisher = {Elsevier},
  doi = {10.1016/B978-0-444-59531-7.00018-1},
  isbn = {1574-0080}
}

@article{odonoghueNoteMethodsMeasuring2004,
  title = {A note on methods for measuring industrial agglomeration},
  author = {O'Donoghue, Dan and Gleave, Bill},
  year = 2004,
  month = jun,
  journal = {Regional Studies},
  volume = {38},
  number = {4},
  pages = {419--427},
  issn = {0034-3404, 1360-0591},
  doi = {10.1080/03434002000213932},
  urldate = {2026-01-11}
}

@article{mcfadden1977modelling,
  title = {Modelling the choice of residential location},
  author = {McFadden, Daniel},
  year = {1977}
}

@article{li2022attenuation,
  title = {Attenuation of agglomeration economies: Evidence from the universe of {Chinese} manufacturing firms},
  author = {Li, Jing and Li, Liyao and Liu, Shimeng},
  journal = {Journal of Urban Economics},
  volume = {130},
  pages = {103458},
  year = {2022},
  publisher = {Elsevier}
}

@article{holmen2022agglomeration,
  title = {Agglomeration decay in rural areas},
  author = {Holmen, Rasmus B{\o}gh},
  journal = {Insights into Regional Development},
  volume = {4},
  number = {3},
  pages = {139--155},
  year = {2022}
}

@article{shannon1948mathematical,
  title = {A mathematical theory of communication},
  author = {Shannon, Claude Elwood},
  journal = {The Bell System Technical Journal},
  volume = {27},
  number = {3},
  pages = {379--423},
  year = {1948},
  publisher = {Nokia Bell Labs}
}

@article{mossel2015reconstruction,
  title = {Reconstruction and estimation in the planted partition model},
  author = {Mossel, Elchanan and Neeman, Joe and Sly, Allan},
  journal = {Probability Theory and Related Fields},
  volume = {162},
  number = {3},
  pages = {431--461},
  year = {2015},
  publisher = {Springer}
}

@article{poudyal2023characterizing,
  title = {Characterizing network circuity among heterogeneous urban amenities},
  author = {Poudyal, Bibandhan and Ghoshal, Gourab and Kirkley, Alec},
  journal = {Journal of the Royal Society Interface},
  volume = {20},
  number = {208},
  year = {2023},
  publisher = {The Royal Society}
}

@article{young2018universality,
  title = {Universality of the stochastic block model},
  author = {Young, Jean-Gabriel and St-Onge, Guillaume and Desrosiers, Patrick and Dub{\'e}, Louis J},
  journal = {Physical Review E},
  volume = {98},
  number = {3},
  pages = {032309},
  year = {2018},
  publisher = {APS}
}

@article{newman2003mixing,
  title = {Mixing patterns in networks},
  author = {Newman, Mark EJ},
  journal = {Physical Review E},
  volume = {67},
  number = {2},
  pages = {026126},
  year = {2003},
  publisher = {APS}
}

@article{holland1983stochastic,
  title = {Stochastic blockmodels: First steps},
  author = {Holland, Paul W and Laskey, Kathryn Blackmond and Leinhardt, Samuel},
  journal = {Social Networks},
  volume = {5},
  number = {2},
  pages = {109--137},
  year = {1983},
  publisher = {Elsevier}
}

@article{openshaw1984modifiable,
  title = {The modifiable areal unit problem},
  author = {Openshaw, Stan},
  journal = {Concepts and Techniques in Modern Geography},
  year = {1984},
  publisher = {GeoBooks}
}

@techreport{overmanCitiesDevelopingWorld2005,
  type = {Working Paper},
  title = {Cities in the Developing World},
  author = {Overman, Henry G. and Venables, Anthony J.},
  year = 2005,
  month = jul,
  number = {695},
  address = {London, UK},
  institution = {{London School of Economics and Political Science. Centre for Economic Performance}},
  urldate = {2026-03-12},
  isbn = {978-0-7530-1875-0}
}

@article{porterClustersNewEconomics1998,
  title = {Clusters and the {{New Economics}} of {{Competition}}},
  author = {Porter, Michael E.},
  year = 1998,
  month = nov,
  journal = {Harvard Business Review},
  issn = {0017-8012},
  urldate = {2026-03-13},
  chapter = {Government policy and regulation}
}

@article{renningerUSCitiesAre2025,
  title = {{{US}} Cities Are Defined by Rings and Pockets with Limited Socioeconomic Mixing},
  author = {Renninger, Andrew and O'Clery, Neave and Arcaute, Elsa},
  year = 2025,
  month = dec,
  journal = {Nature Cities},
  volume = {2},
  number = {12},
  pages = {1172--1182},
  issn = {2731-9997},
  doi = {10.1038/s44284-025-00350-7},
  urldate = {2026-01-11}
}

@incollection{rosenthalChapter49Evidence2004,
  title = {{Chapter 49 - Evidence on the Nature and Sources of Agglomeration Economies}},
  booktitle = {Cities and Geography},
  author = {Rosenthal, Stuart S. and Strange, William C.},
  editor = {Henderson, J. Vernon and Thisse, Jacques-Fran{\c c}ois},
  year = 2004,
  series = {Handbook of Regional and Urban Economics},
  volume = {4},
  pages = {2119--2171},
  publisher = {Elsevier},
  issn = {1574-0080},
  doi = {10.1016/S1574-0080(04)80006-3}
}

@article{schlapferUniversalVisitationLaw2021,
  title = {The Universal Visitation Law of Human Mobility},
  author = {Schl{\"a}pfer, Markus and Dong, Lei and O'Keeffe, Kevin and Santi, Paolo and Szell, Michael and Salat, Hadrien and Anklesaria, Samuel and Vazifeh, Mohammad and Ratti, Carlo and West, Geoffrey B.},
  year = 2021,
  month = may,
  journal = {Nature},
  volume = {593},
  number = {7860},
  pages = {522--527},
  publisher = {Nature Publishing Group},
  issn = {1476-4687},
  doi = {10.1038/s41586-021-03480-9},
  urldate = {2026-03-14}
}

@article{storperBuzzFacetofaceContact2004,
  title = {Buzz: Face-to-Face Contact and the Urban Economy},
  shorttitle = {Buzz},
  author = {Storper, Michael and Venables, Anthony J.},
  year = 2004,
  month = aug,
  journal = {Journal of Economic Geography},
  volume = {4},
  number = {4},
  pages = {351--370},
  issn = {1468-2702},
  doi = {10.1093/jnlecg/lbh027},
  urldate = {2026-03-12}
}

@book{westScaleUniversalLaws2017,
  title = {Scale: {{The Universal Laws}} of {{Growth}}, {{Innovation}}, {{Sustainability}}, and the {{Pace}} of {{Life}} in {{Organisms}}, {{Cities}}, {{Economies}}, and {{Companies}}},
  shorttitle = {Scale},
  author = {West, Geoffrey},
  year = 2017,
  month = apr,
  publisher = {Penguin Group , The},
  isbn = {978-1-59420-558-3}
}

@book{jacobsDeathLifeGreat1961,
  title = {The Death and Life of Great {American} Cities},
  author = {Jacobs, Jane},
  year = 1961,
  publisher = {Random House},
  address = {New York}
}

@article{delclsali2019urban,
  title = {The urban vitality conditions of {Jane Jacobs} in {Barcelona}: Residential and smartphone-based tracking measurements of the built environment in a {Mediterranean} metropolis},
  author = {Xavier Delclòs-Alió and Aaron Gutiérrez and Carme Miralles-Guasch},
  year = {2019},
  publisher = {Elsevier BV},
  journal = {Cities},
  volume = {86},
  pages = {220-228},
  doi = {10.1016/j.cities.2018.09.021},
}

@article{mouratidis2020built,
  title = {Built environment, urban vitality and social cohesion: Do vibrant neighborhoods foster strong communities?},
  author = {Kostas Mouratidis and Wouter Poortinga},
  year = {2020},
  publisher = {Elsevier BV},
  journal = {Landscape and Urban Planning},
  volume = {204},
  pages = {103951},
  doi = {10.1016/j.landurbplan.2020.103951},
}

@misc{hkfehdRestaurantLicences2026,
  title = {Restaurant Licences ({English})},
  author = {{Food and Environmental Hygiene Department}},
  year = {2026},
  howpublished = {\url{https://data.gov.hk/en-data/dataset/hk-fehd-fehdlmis-restaurant-licences}}
}

@misc{hkcensusLTPUG2016,
  title = {2016 Population By-census (Statistics and Boundaries of Large Tertiary Planning Unit Groups)},
  author = {{Census and Statistics Department}},
  year = {2026},
  howpublished = {\url{https://data.gov.hk/en-data/dataset/hk-censtatd-census\_geo-2016-population-bycensus-by-ltpu}}
}

@misc{hkcensusLTPUG2021,
  title = {2021 Population Census (Statistics and Boundaries of Large Tertiary Planning Unit Groups)},
  author = {{Census and Statistics Department}},
  year = {2026},
  howpublished = {\url{https://data.gov.hk/en-data/dataset/hk-censtatd-census\_geo-2021-population-census-by-ltpu}}
}

@misc{googleGeocodingAPI2026,
  title = {Geocoding {API} v4 Overview},
  author = {{Google}},
  url = {https://developers.google.com/maps/documentation/geocoding/geocoding-v4-overview},
  year = {2026}
}

@misc{googlePlaceAPI2026,
  title = {Places {API}},
  author = {{Google}},
  year = {2026},
  howpublished = {\url{https://developers.google.com/maps/documentation/places/web-service/place-types}}
}

@misc{gemini2025,
  title = {{Gemini} 2.5: Pushing the Frontier with Advanced Reasoning, Multimodality, Long Context, and Next Generation Agentic Capabilities},
  author = {Google DeepMind},
  year = {2025},
  eprint = {2507.06261},
  archiveprefix = {arXiv},
  primaryclass = {cs.CL},
  url = {https://arxiv.org/abs/2507.06261}
}

@article{kaufmannScalingUrbanAmenities2022,
  title = {Scaling of Urban Amenities: Generative Statistics and Implications for Urban Planning},
  shorttitle = {Scaling of Urban Amenities},
  author = {Kaufmann, Talia and Radaelli, Laura and Bettencourt, Luis M. A. and Shmueli, Erez},
  year = 2022,
  month = sep,
  journal = {EPJ Data Science},
  volume = {11},
  number = {1},
  pages = {50},
  issn = {2193-1127},
  doi = {10.1140/epjds/s13688-022-00362-6},
  urldate = {2026-03-30}
}

@misc{foursquare,
  author = {{Foursquare Labs, Inc.}},
  title = {Open Source {POI} Data | {Foursquare OS} Places},
  year = {2026},
  howpublished = {\url{https://opensource.foursquare.com/os-places/}}
}

@misc{wipfliWipfliFoursquareosplacespmtiles2026,
  title = {Wipfli/Foursquare-Os-Places-Pmtiles},
  author = {Wipfli, Oliver},
  year = 2026,
  month = mar,
  urldate = {2026-04-01}
}

@misc{esriWorldUrbanAreas,
  author = {{Esri}},
  title = {World Urban Areas},
  year = {2025},
  month = apr,
  howpublished = {\url{https://hub.arcgis.com/datasets/esri::world-urban-areas}},
  urldate = {2026-04-01}
}

@article{kuang2017does,
  title = {Does quality matter in local consumption amenities? An empirical investigation with {Yelp}},
  author = {Chun Kuang},
  year = {2017},
  publisher = {Elsevier BV},
  journal = {Journal of Urban Economics},
  volume = {100},
  pages = {1-18},
  doi = {10.1016/j.jue.2017.02.006},
}

@article{su2022measuring,
  title = {Measuring the Value of Urban Consumption Amenities: A Time-Use Approach},
  author = {Yichen Su},
  year = {2022},
  publisher = {Elsevier BV},
  journal = {Journal of Urban Economics},
  volume = {132},
  pages = {103495},
  doi = {10.1016/j.jue.2022.103495},
}

@article{sui2024economic,
  title = {Economic value of 10-min neighborhood: Evidence from {Munich} Metropolitan Area, {Germany}},
  author = {Yijun Sui and Bing Zhu},
  year = {2024},
  publisher = {Elsevier BV},
  journal = {Cities},
  volume = {154},
  pages = {105370},
  doi = {10.1016/j.cities.2024.105370},
}

@article{glaesener2015neighborhood,
  title = {Neighborhood green and services diversity effects on land prices: Evidence from a multilevel hedonic analysis in {Luxembourg}},
  author = {Marie-Line Glaesener and Geoffrey Caruso},
  year = {2015},
  publisher = {Elsevier BV},
  journal = {Landscape and Urban Planning},
  volume = {143},
  pages = {100-111},
  doi = {10.1016/j.landurbplan.2015.06.008},
}

@article{heine2025role,
  title = {The role of urban amenities in facilitating social mixing: Evidence from {Stockholm}},
  author = {Cate Heine and Timur Abbiasov and Paolo Santi and Carlo Ratti},
  year = {2025},
  publisher = {Elsevier BV},
  journal = {Landscape and Urban Planning},
  volume = {254},
  pages = {105250},
  doi = {10.1016/j.landurbplan.2024.105250},
}

@article{liu2026agglomeration,
  author = {Liu, Crocker H. and Zheng, Chen and Zhu, Bing},
  month = {01},
  publisher = {Institute for Operations Research and the Management Sciences (INFORMS)},
  title = {Does Agglomeration Enhance Property Value?},
  doi = {10.1287/mnsc.2023.02773},
  year = {2026},
  journal = {Management Science}
}

@article{liu2024agglomeration,
  title = {Agglomeration Economies and the Built Environment: Evidence from Specialized Buildings and Anchor Tenants},
  author = {Crocker H. Liu and Stuart S. Rosenthal and William C. Strange},
  year = {2024},
  publisher = {Elsevier BV},
  journal = {Journal of Urban Economics},
  volume = {142},
  pages = {103655},
  doi = {10.1016/j.jue.2024.103655},
}

@article{yu2026characterizing,
  title = {Characterizing walkability in {Hong Kong}'s 15-minute transit-oriented development ({TOD}): insights from street view imagery and local accessibility},
  author = {Zidong Yu and Ketong Shen and Xintao Liu},
  year = {2026},
  publisher = {Elsevier BV},
  journal = {Travel Behaviour and Society},
  volume = {42},
  pages = {101157},
  doi = {10.1016/j.tbs.2025.101157},
}

@article{miao2025vibrancy,
  title = {Vibrancy of the space around and between metro stations: Life between Stations and its determinants in {Hong Kong}},
  author = {Yunting Miao and Jiangping Zhou},
  year = {2025},
  publisher = {Springer Science and Business Media LLC},
  journal = {Transportation},
  doi = {10.1007/s11116-025-10710-w},
}

\clearpage
\onecolumngrid

\setcounter{equation}{0}
\setcounter{figure}{0}
\setcounter{table}{0}
\setcounter{page}{1}
\setcounter{section}{0}
\renewcommand{\theequation}{A\arabic{equation}}
\renewcommand{\thefigure}{A\arabic{figure}}
\renewcommand{\thetable}{A\arabic{table}}
\renewcommand{\thepage}{A\arabic{page}}
\renewcommand{\thesection}{A\arabic{section}}
\renewcommand{\theHequation}{A\arabic{equation}}
\renewcommand{\theHfigure}{A\arabic{figure}}
\renewcommand{\theHtable}{A\arabic{table}}
\renewcommand{\theHsection}{A\arabic{section}}
\renewcommand{\bibnumfmt}[1]{[#1]}
\renewcommand{\citenumfont}[1]{#1}

\section{Description length expressions}
\label{sec:dl-derivations}

In this appendix we derive the expressions used to evaluate the description lengths in Eqs.~\eqref{eq:dl-hom},~\eqref{eq:dl-het}~and~\eqref{eq:dl-neu} for a given network $\bm{A}$ with node labels $\bm{b}$. 

\paragraph{Homophilic model.}
Using Eqs. \eqref{eq:prior-hom} and \eqref{eq:marginal-lik}, the homophilic marginal likelihood becomes
\begin{align}
P_{ml}^{(hom)}(\bm{A},\bm{b})
&=2P_{part}(\bm{b})\int_{0}^{1}\int_{0}^{1} P_{ppm}(\bm{A}\vert \bm{b},p_{in},p_{out}) \Theta(p_{in}-p_{out})dp_{in}dp_{out}
\nonumber\\
&=
2P_{part}(\bm{b})
\int_0^1\int_0^{p_{\mathrm{in}}}
p_{\mathrm{in}}^{E_{\mathrm{in}}}(1-p_{\mathrm{in}})^{Q_{\mathrm{in}}-E_{\mathrm{in}}}
p_{\mathrm{out}}^{E_{\mathrm{out}}}(1-p_{\mathrm{out}})^{Q_{\mathrm{out}}-E_{\mathrm{out}}}
dp_{\mathrm{out}}dp_{\mathrm{in}}
\nonumber\\
&=
2P_{part}(\bm{b})
\int_0^1
p_{\mathrm{in}}^{E_{\mathrm{in}}}(1-p_{\mathrm{in}})^{Q_{\mathrm{in}}-E_{\mathrm{in}}}
\mathrm{B}(p_{\mathrm{in}};E_{\mathrm{out}}+1,Q_{\mathrm{out}}-E_{\mathrm{out}}+1)
dp_{\mathrm{in}},
\end{align}
where $\mathrm{B}(x;a,b)=\int_0^x t^{a-1}(1-t)^{b-1}dt$ is the incomplete beta function. Therefore
\begin{align}
\mathcal{L}_{hom}
=
\mathcal{L}_{\bm{b}}
-\log 2
-\log \int_0^1
p_{\mathrm{in}}^{E_{\mathrm{in}}}(1-p_{\mathrm{in}})^{Q_{\mathrm{in}}-E_{\mathrm{in}}}
\mathrm{B}(p_{\mathrm{in}};E_{\mathrm{out}}+1,Q_{\mathrm{out}}-E_{\mathrm{out}}+1)
dp_{\mathrm{in}},
\label{eq:lhom-int}
\end{align}
in which the description length of the prior on the partition $\bm{b}$, by Eq. \eqref{eq:prior-b}, is
\begin{align}
\mathcal{L}_{\bm{b}}=-\log P_{part}(\bm{b})
=
\log N+\log\binom{N-1}{B-1}+\log\binom{N}{n_1,\dots,n_B}.
\label{eq:dl-b}
\end{align}

\paragraph{Heterophilic model.}
Likewise, using Eqs. \eqref{eq:prior-het} and \eqref{eq:marginal-lik} gives
\begin{align}
P_{ml}^{(het)}(\bm{A},\bm{b})
&=2P_{part}(\bm{b})\int_{0}^{1}\int_{0}^{1} P_{ppm}(\bm{A}\vert \bm{b},p_{in},p_{out}) \Theta(p_{out}-p_{in})dp_{in}dp_{out}
\nonumber\\
&=
2P_{part}(\bm{b})
\int_0^1\int_0^{p_{\mathrm{out}}}
p_{\mathrm{in}}^{E_{\mathrm{in}}}(1-p_{\mathrm{in}})^{Q_{\mathrm{in}}-E_{\mathrm{in}}}
p_{\mathrm{out}}^{E_{\mathrm{out}}}(1-p_{\mathrm{out}})^{Q_{\mathrm{out}}-E_{\mathrm{out}}}
dp_{\mathrm{in}}dp_{\mathrm{out}}
\nonumber\\
&=
2P_{part}(\bm{b})
\int_0^1
p_{\mathrm{out}}^{E_{\mathrm{out}}}(1-p_{\mathrm{out}})^{Q_{\mathrm{out}}-E_{\mathrm{out}}}
\mathrm{B}(p_{\mathrm{out}};E_{\mathrm{in}}+1,Q_{\mathrm{in}}-E_{\mathrm{in}}+1)
dp_{\mathrm{out}},
\end{align}
and hence
\begin{align}
\mathcal{L}_{het}
=
\mathcal{L}_{\bm{b}}
-\log 2
-\log \int_0^1
p_{\mathrm{out}}^{E_{\mathrm{out}}}(1-p_{\mathrm{out}})^{Q_{\mathrm{out}}-E_{\mathrm{out}}}
\mathrm{B}(p_{\mathrm{out}};E_{\mathrm{in}}+1,Q_{\mathrm{in}}-E_{\mathrm{in}}+1)
dp_{\mathrm{out}}.
\label{eq:lhet-int}
\end{align}

\paragraph{Neutral model.}
Under the neutral model, the partition labels are irrelevant. Using Eqs. \eqref{eq:prior-neu} and \eqref{eq:marginal-lik}, we have
\begin{align}
P_{ml}(\bm{A}) 
&=\int_{0}^{1}\int_{0}^{1} P_{ppm}(\bm{A}\vert p_{in},p_{out})\delta(p_{in}-p_{out})dp_{in}dp_{out}
\nonumber\\
&=\int_{0}^{1} P_{ppm}(\bm{A}\vert p,p)dp
\nonumber\\
&=\int_0^1 p^E(1-p)^{Q-E}dp
\nonumber\\
&=
\mathrm{B}(E+1,Q-E+1),
\end{align}
where $\mathrm{B}(a,b)$ is the beta function. Using the factorial form of the beta function gives
\begin{align}
P_{ml}^{(neu)}(\bm{A})
&=
\frac{E!(Q-E)!}{(Q+1)!}
=
\frac{1}{(Q+1)\binom{Q}{E}}.
\label{eq:pneu-comb}
\end{align}
Hence the neutral description length is
\begin{align}
\mathcal{L}_{neu}=-\log P_{ml}^{(neu)}(\bm{A}) =\log(Q+1)+\log\binom{Q}{E}.
\label{eq:lneu-comb}
\end{align}
Equation~\eqref{eq:lneu-comb} can be interpreted as the information required to first specify the total number of edges $E$ uniformly from $\{0,1,\dots,Q\}$, plus the information required to specify which $E$ of the $Q$ possible edges are present in $\bm{A}$.

\section{Algorithmic implementation}
\label{sec:algorithm}

We now describe how the description lengths are evaluated numerically for a point set $\bm{X}$ of $N$ amenities over a grid of distance thresholds, and analyze the computational cost of the procedure. Suppose that the description lengths are to be evaluated on an ordered grid of $K$ distance thresholds $0 < \epsilon_1 < \epsilon_2 < \cdots < \epsilon_K$. In practice one may evaluate $T$ partitions of the same point set simultaneously into different amenity types, either because amenities are available at different levels of categorization or because one wishes to isolate a single category against all others and thereby form a sequence of binary labels (as done in Fig.~\ref{fig:results_foursquare}). For each partition $\bm{b}^{(t)},~t=1,...,T$, the only scale-dependent quantities required by Appendix~\ref{sec:dl-derivations} are $E(\epsilon_k)$ and $E_{in}(\epsilon_k)$. $E_{out}(\epsilon_k)=E-E_{in}(\epsilon_k)$ is determined by these two quantities, and $Q$, $Q_{in}$, and $Q_{out}$ can be computed immediately by the counts of each amenity type in $\bm{b}^{(t)}$.

For algorithmic efficiency, we avoid rebuilding the proximity graph from scratch at every threshold. Instead, pairwise distances are computed once and histogrammed into the $K$ distance bins $[0,\epsilon_1],\,(\epsilon_1,\epsilon_2],\,\dots,\,(\epsilon_{K-1},\epsilon_K]$. For every unordered pair $(i,j)$, the distance $d_{ij}=d(\bm{x}_i,\bm{x}_j)$ is assigned to its bin, incrementing both a global histogram over all pairs and a within-group histogram for each of the $T$ partitions being tested. Once all pairs have been processed, cumulative sums of these histograms yield $E(\epsilon_k)$ and $E_{in}(\epsilon_k)$ for every threshold $\epsilon_k$ simultaneously.

Because storing all $\binom{N}{2}$ pairwise distances is prohibitive for large $N$, the point set is divided into chunks $u,u',\ldots$ of size at most $c$. The algorithm loops over chunk pairs $(u,u')$ with $u\leq u'$: when $u=u'$, only the upper-triangular within-chunk distances are computed; when $u\neq u'$, the full cross-distance matrix between the two chunks is formed. In either case, the resulting distances are histogrammed into the $K$ bins to update the total and within-group edge counts. All of these operations---distance computation, masking, and histogramming---are vectorized at the block level, so the work is dominated by dense array arithmetic rather than Python loops over individual point pairs. For the same reason, the standard choice of KD-tree for such spatial indexing is not advantageous when the largest threshold is large enough that most points interact with a substantial fraction of the dataset, which is the regime relevant here. The chunked dense computation is both simpler and faster for this case. 

Once $E(\epsilon_k)$ and $E_{in}(\epsilon_k)$ are available, the neutral description length is obtained in closed form from Eq.~\eqref{eq:lneu-comb}. The homophilic and heterophilic description lengths require numerical integration of the one-dimensional expressions in Eqs.~\eqref{eq:lhom-int} and~\eqref{eq:lhet-int}, which can be easily evaluated with Gauss--Legendre quadrature vectorized across all partitions and thresholds via array broadcasting. The partition description length $\mathcal{L}_{\bm{b}}$ is then added to the conditional description lengths to produce the full model description lengths.

The overall computational cost can be estimated as follows. The distance computation and histogramming step visits every chunk pair $(u,u')$ with $u \leq u'$, of which there are $\lceil N/c \rceil (\lceil N/c \rceil + 1)/2$. Each pair requires $O(c^2)$ computations for distances and histogramming, and a further $O(c^2 T)$ work for the $T$ within-group masks and histograms, giving a total of $O(N^2 T)$ for the sweep if $\epsilon_K$ is set to its largest possible value and we explore $T$ distinct partitions of the points. The cumulative sums over bins require $O(TK)$ steps, while the Gauss--Legendre quadrature evaluates the integrand at a specific number of points $M$ for each of the $TK$ (partition, threshold) combinations, for a complexity of $O(MTK)$. The partition prior $\mathcal{L}_{\bm{b}}$ requires $O(N)$ steps per partition, which is negligible in comparison. Altogether the running time for scanning over all point pairs is $O\!\left(N^2 T + MTK\right)$, with $O(N^2 T)$ the dominating complexity scaling in practice. The peak memory scales as $O(c^2 +TN+ TK)$, with $O(c^2)$ for the distance matrix of a single chunk pair, $O(TN)$ for label arrays, and $O(TK)$ for the histograms. Thus the chunk size $c$ can be tuned to trade wall-clock time for memory.

The algorithm returns, for every node partition $\bm{b}^{(t)}$ and every distance threshold $\epsilon_k$, the arrays $\mathcal{L}_{neu}(\epsilon_k)$, $\mathcal{L}_{hom}(\epsilon_k)$, and $\mathcal{L}_{het}(\epsilon_k)$, together with the intermediate counts $(E,E_{in},Q,Q_{in})$. These trajectories can then be used directly for model selection, for identifying the characteristic scales $\epsilon_{hom}^\ast$ and $\epsilon_{het}^\ast$, and for computing compression ratios as in Eq.~\eqref{eq:ratio}.

\section{Phase transition between homophilic and heterophilic regimes}
\label{sec:analytical-properties}

To understand how this MDL framework will categorize spatial amenity configurations in practice, we first perform asymptotic analysis on the marginal likelihoods to identify the regimes in which the description length expressions in Eqs.~\ref{eq:dl-hom},~\ref{eq:dl-het},~and~\ref{eq:dl-neu} will be minimal.

Using Eq.~\ref{eq:marginal-lik} and performing a Laplace approximation, we have
\begin{align}
P_{ml}(\bm{A},\bm{b}) &= P_{part}(\bm{b})\int_{0}^{1}\int_{0}^{1} P_{ppm}(\bm{A}\vert \bm{b},p_{in},p_{out}) P_{mix}(p_{in},p_{out})dp_{in}dp_{out} \\
&\approx P_{part}(\bm{b}) \times \frac{2\pi}{\sqrt{\det H}} \times P_{ppm}(\bm{A}\vert \bm{b},\hat{p}_{in},\hat{p}_{out}) P_{mix}(\hat{p}_{in},\hat{p}_{out}) ,
\end{align}
where $H$ is the Hessian matrix of $-\log P_{ppm}$ evaluated at the maximum likelihood estimates $\hat{p}_{in}=E_{in}/Q_{in}$, $\hat{p}_{out}=E_{out}/Q_{out}$, provided these estimates lie in the interior of the support of $P_{mix}$. For the homophilic model, this requires $\hat{p}_{in} \geq \hat{p}_{out}$, and for the heterophilic model, this requires $\hat{p}_{in} \leq \hat{p}_{out}$. The MLEs $\hat{p}_{in}$ and $\hat{p}_{out}$ have a natural interpretation as the observed density of edges among nodes sharing the same label and the observed density of edges between nodes with different labels, respectively.

Taking the negative logarithm to obtain the asymptotic description length,
\begin{align}
\mathcal{L}_{\bm{X},\bm{b}}(\epsilon) &= -\log P_{ml}(\bm{A},\bm{b})\\
&\approx -\log P_{part}(\bm{b}) - \log\left(\frac{2\pi}{\sqrt{\det H}}\right) -\log P_{ppm}(\bm{A}\vert \bm{b},\hat{p}_{in},\hat{p}_{out}) - \log P_{mix}(\hat{p}_{in},\hat{p}_{out}).
\end{align}
The dominant contribution comes from $-\log P_{ppm}$, which scales as $O(N^2)$ for $E\sim O(N^2)$ or $O(N\log N)$ for $E\sim O(N)$ and is identical for both the homophilic and heterophilic models. The remaining terms scale as $-\log P_{part}(\bm{b}) \sim O(N\log B)\sim O(N)$ for a constant number of categories $B$, $-\log(2\pi/\sqrt{\det H}) \sim O(\log N)$, and $-\log P_{mix}(\hat{p}_{in},\hat{p}_{out}) \sim O(1)$. Thus, the likelihood term $-\log P_{ppm}$ provides the dominant contribution to the description length, and the transition between the homophilic and heterophilic regimes is determined by whether the MLEs $\hat p_{in},\hat p_{out}$ satisfy the prior constraints of $P_{mix}$. If $\hat p_{in},\hat p_{out}$ are within the support of $P_{mix}$, the description length is asymptotically dominated by $-\log P_{ppm}(\bm{A}\vert \bm{b},\hat p_{in},\hat p_{out})$, and is of lower order in $N$ if not.

This analysis thus reveals a sharp phase transition (as $N,E\to\infty$) between the homophilic and heterophilic regimes at the spatial scales $\epsilon$ for which
\begin{align}
\hat{p}_{in}(\epsilon) = \hat{p}_{out}(\epsilon) \quad \Longleftrightarrow \quad \frac{E_{in}(\epsilon)}{Q_{in}} = \frac{E_{out}(\epsilon)}{Q_{out}}.
\end{align}
In other words, when we are at distance scales for which the density of in-edges $\hat{p}_{in}(\epsilon)$ is greater than the density of out-edges $\hat{p}_{out}(\epsilon)$, the amenity configuration will be classified as homophilic since $\mathcal{L}^{(hom)}_{\bm{X},\bm{b}}(\epsilon)<\mathcal{L}^{(het)}_{\bm{X},\bm{b}}(\epsilon)$. And when we are at distance scales for which the density of in-edges $\hat{p}_{in}(\epsilon)$ is less than the density of out-edges $\hat{p}_{out}(\epsilon)$, the amenity configuration will be classified as heterophilic since $\mathcal{L}^{(hom)}_{\bm{X},\bm{b}}(\epsilon)>\mathcal{L}^{(het)}_{\bm{X},\bm{b}}(\epsilon)$. At the critical point $\hat{p}_{in}(\epsilon)=\hat{p}_{out}(\epsilon)$, both homophilic and heterophilic models yield the same asymptotic marginal likelihood, since the MLEs $\hat{p}_{in}(\epsilon),\hat{p}_{out}(\epsilon)$ are within the support of both models' priors $P_{mix}$.

For finite networks outside of the asymptotic regime, the non-informative prior on the partition $\bm{b}$ introduces a penalty $-\log P_{part}(\bm{b})\geq 0$ that favors the neutral model, since the neutral model does not require this specification of $\bm{b}$ to transmit the network structure. This effect is most pronounced in small or sparse graphs (i.e. small distance scales), where the likelihood term $-\log P_{ppm}$ is comparatively weak, creating a neutral region around the $\hat p_{in} = \hat p_{out}$ transition line where neither structured model is decisively favored.

\section{Physically consistent generative model for urban amenities}
\label{sec:physical}

The key advantage of SBMs as a modeling framework for spatial amenity data is that they provide analytically tractable probabilistic expressions that permit rigorous Bayesian model selection to identify the homophilic, heterophilic, and neutral mixing regimes. SBMs do not, however, provide a truly plausible physical generative process for spatial data, since the independence assumption among the edges in the generation process is broken by the triangle inequality in metric spaces. In other words, if amenity a and amenity b share an edge due to proximity, and amenity a and amenity c share an edge due to proximity, then it is likely that amenity b and amenity c share an edge due to proximity, so we cannot say that the edge $(b,c)$ is generated independently of $(a,b)$ and $(a,c)$. (The reverse holds true if edges are absent between the node pairs.) This suggests that a more physically plausible spatial model, while analytically intractable, can provide a complementary perspective of value for interpreting the findings of the SBM framework.

Consider a set of $N$ amenities with fixed categories $\bm{b}$ as before. However, now allow the spatial positions $\bm{X}$ to be dynamic, reflecting the ability for the amenity locations to be chosen strategically. Let the amenities move within a fixed two-dimensional spatial domain $\Omega$ of area $V$, representing the natural boundaries (e.g. city or neighborhood within the city) where the amenities may locate. The spatial domain is assumed to be relatively convex and much larger than the size of each amenity, so its exact shape does not impact the model dynamics. Also assume that there is a minimal distance $R_{min}$ between plots on which the amenities will locate, such that $\vert\vert \bm{x}_i-\bm{x}_j\vert\vert \geq R_{min}$ for all amenities $i,j$ (i.e., amenities cannot overlap in space).

We can then consider a pairwise utility function $U_{ij}(r)$ that represents the agglomerative/co-agglomerative benefits the amenities $i$ and $j$ each receive by being at distance $r$ from each other. A simple form for this utility function is
\begin{align}
U_{ij}(r) = 
\begin{cases}
-\infty, & r \leq R_{min}, \\
J_{b_ib_j}\phi(r/\xi), & r > R_{min}
\end{cases}
\end{align}
where $J_{b_ib_j}$ determines the sign and magnitude of benefits incurred from co-location, while $\phi(r/\xi)$ is a strictly decreasing function that determines the attenuation of agglomerative benefits and costs with spatial distance $r$. The utility is set to $U=-\infty$ for $r \leq R_{min}$ to impose a hard constraint that prevents amenities from overlapping. The parameter $\xi$ is a characteristic spatial scale associated with the attenuation process. Based on previous empirical evidence, the functional forms $\phi(r)=e^{-r/\xi}$ \cite{holmen2022agglomeration} and $\phi(r)=(r/\xi)^{-\alpha}$ \cite{li2022attenuation} provide decay behavior consistent with different firms and amenities. (Specifically, the inverse-square form $\phi(r)=r^{-2}$ was found to be consistent with the agglomeration tendencies of Chinese manufacturing firms \cite{li2022attenuation}.)

The total utility for all firms/amenities when they have spatial configuration $\bm{X}$ is then given by
\begin{align}
U_{tot}(\bm{X} \vert \bm{b}) = \sum_{i<j} U_{ij}(\|\bm{x}_i - \bm{x}_j\|),
\end{align}
which holds for configurations $\bm{X}$ within $\Omega$. Following the framework of random utility models \cite{mcfadden1977modelling} and spin models in statistical physics, we can model the probability of observing a particular configuration $\bm{X}$ of the amenities using the Boltzmann distribution
\begin{align}
P_{conf}(\bm{X} \vert \bm{b}) \propto e^{U_{tot}(\bm{X} \vert \bm{b})} = \prod_{i<j}e^{U_{ij}(\|\bm{x}_i - \bm{x}_j\|)},
\end{align}
where we have omitted the normalization constant for brevity. This spatial amenity location process induces a marginal distribution over proximity networks $\bm{A}$ for distance scale $\epsilon$, given by
\begin{align}
P_{conf}(\bm{A} \vert \bm{b}) = \int_{\bm{X}: \bm{A}(\bm{X},\epsilon)=\bm{A}} P_{conf}(\bm{X} \vert \bm{b}) d\bm{X}
\propto \int_{\bm{X}}  \prod_{i<j}e^{U_{ij}(\|\bm{x}_i - \bm{x}_j\|)} g(\vert\vert \bm{x}_i-\bm{x}_j\vert\vert,A_{ij},\epsilon)d\bm{X},
\end{align}
where
\begin{align}
g(\vert\vert \bm{x}_i-\bm{x}_j\vert\vert,A_{ij},\epsilon) = \Theta(\epsilon-\vert\vert \bm{x}_i-\bm{x}_j\vert\vert)^{A_{ij}}\Theta(\vert\vert \bm{x}_i-\bm{x}_j\vert\vert-\epsilon)^{1-A_{ij}}    
\end{align}
enforces the matching $\bm{A}=\bm{A}(\bm{X},\epsilon)$. This integral is in general intractable, but in the regime of low interaction density $N\epsilon^2/V\ll 1$ and small agglomeration utility $\abs{U_{ij}}\ll 1$, we can apply a mean-field pair approximation, giving
\begin{align}
P_{conf}(\bm{A} \vert \bm{b})\propto  \prod_{i<j}\int e^{U_{ij}(\|\bm{x}_i - \bm{x}_j\|)} g(\vert\vert \bm{x}_i-\bm{x}_j\vert\vert,A_{ij},\epsilon)d\bm{x}_i d\bm{x}_j \int d\bm{X}^{-\{i,j\}}
\propto \prod_{i<j}\int e^{U_{ij}(\|\bm{x}_i - \bm{x}_j\|)} g(\vert\vert \bm{x}_i-\bm{x}_j\vert\vert,A_{ij},\epsilon)d\bm{x}_i d\bm{x}_j, 
\end{align}
where $\bm{X}^{-\{i,j\}}$ is the set of coordinates excluding $\bm{x}_i,\bm{x}_j$.

Now, if the agglomeration utility $U_{ij}(r)$ only depends on the distance $r$ and whether or not the amenities are of the same or different categories, we can set
\begin{align}
J_{b_ib_j}=J_{in}\delta_{b_ib_j}+J_{out}(1-\delta_{b_ib_j}),    
\end{align}
which simplifies the marginal likelihood to
\begin{align}
P_{conf}(\bm{A} \vert \bm{b}) &\propto \left[\int e^{J_{in}\phi(\|\bm{x}_i - \bm{x}_j\|/\xi)} \Theta(\epsilon-\vert\vert \bm{x}_i-\bm{x}_j\vert\vert)d\bm{x}_i\bm{x}_j\right]^{E_{in}}\times \left[\int e^{J_{in}\phi(\|\bm{x}_i - \bm{x}_j\|/\xi)} \Theta(\vert\vert \bm{x}_i-\bm{x}_j\vert\vert-\epsilon)d\bm{x}_i\bm{x}_j\right]^{Q_{in}-E_{in}} \nonumber\\
&\times \left[\int e^{J_{out}\phi(\|\bm{x}_i - \bm{x}_j\|/\xi)} \Theta(\epsilon-\vert\vert \bm{x}_i-\bm{x}_j\vert\vert)d\bm{x}_i\bm{x}_j\right]^{E_{out}}\times \left[\int e^{J_{out}\phi(\|\bm{x}_i - \bm{x}_j\|/\xi)} \Theta(\vert\vert \bm{x}_i-\bm{x}_j\vert\vert-\epsilon)d\bm{x}_i\bm{x}_j\right]^{Q_{out}-E_{out}} \\
&\propto \left[\int_{R_{min}}^{\epsilon} r e^{J_{in}\phi(r/\xi)} dr\right]^{E_{in}}\times \left[\int_{\epsilon}^{R_V} r e^{J_{in}\phi(r/\xi)} dr\right]^{Q_{in}-E_{in}} \nonumber\\
&\times \left[\int_{R_{min}}^{\epsilon} r e^{J_{out}\phi(r/\xi)} dr\right]^{E_{out}}\times \left[\int_{\epsilon}^{R_V} r e^{J_{out}\phi(r/\xi)} dr\right]^{Q_{out}-E_{out}},
\end{align}
where we have switched to polar coordinates and assumed translational invariance (i.e. periodic boundary conditions). The upper cutoff $R_V$ is an effective finite radial cutoff determined by the spatial domain; for example, in the disk approximation $R_V=\sqrt{V/\pi}$. We can see that this marginal likelihood is just proportional to
\begin{align}
P_{conf}(\bm{A} \vert \bm{b}) \propto p_{in}(\xi,\epsilon)^{E_{in}}[1-p_{in}(\xi,\epsilon)]^{Q_{in}-E_{in}}    
p_{out}(\xi,\epsilon)^{E_{out}}[1-p_{out}(\xi,\epsilon)]^{Q_{out}-E_{out}},    
\end{align}
where
\begin{align}
p_{in}(\xi,\epsilon) &= \frac{\int_{R_{min}}^{\epsilon} r e^{J_{in}\phi(r/\xi)} dr}{\int_{R_{min}}^{R_V} r e^{J_{in}\phi(r/\xi)} dr} \\
p_{out}(\xi,\epsilon) &= \frac{\int_{R_{min}}^{\epsilon} r e^{J_{out}\phi(r/\xi)} dr}{\int_{R_{min}}^{R_V} r e^{J_{out}\phi(r/\xi)} dr} \\
\end{align}
are the probabilities for a pair of points to fall within the distance threshold $\epsilon$ given that they are the same amenity type or of different amenity types, respectively. We can now see a direct correspondence between the marginal likelihood of the proximity network $\bm{A}$ under a spatial random utility process with agglomeration benefits and the likelihood of the proximity network $\bm{A}$ in the planted partition model. This provides a mapping between the purely relational network model of the SBM and a physical generative model of amenity location choice.

The MDL classification in the main text selects the model (homophilic, heterophilic, or neutral) that minimizes the description length. The line $\hat{p}_{in} = \hat{p}_{out}$ separating the homophilic and heterophilic regimes then corresponds precisely to $J_{in} = J_{out}$ in this spatial utility model. Homophilic ordering ($J_{in} > J_{out}$) corresponds to stronger agglomerative benefits between same-type amenities than between different-type amenities, while heterophilic ordering ($J_{out} > J_{in}$) corresponds to stronger agglomerative benefits between different-type amenities. 

\clearpage
\section{Supplementary Foursquare Analyses}
\label{sec:supp-foursquare}
\vspace{-10pt}
\begingroup
\setlength{\intextsep}{0.35\baselineskip}
\setlength{\abovecaptionskip}{0.25\baselineskip}
\setlength{\belowcaptionskip}{0pt}

\begin{figure}[H]
	\centering
	\includegraphics[width=1\linewidth]{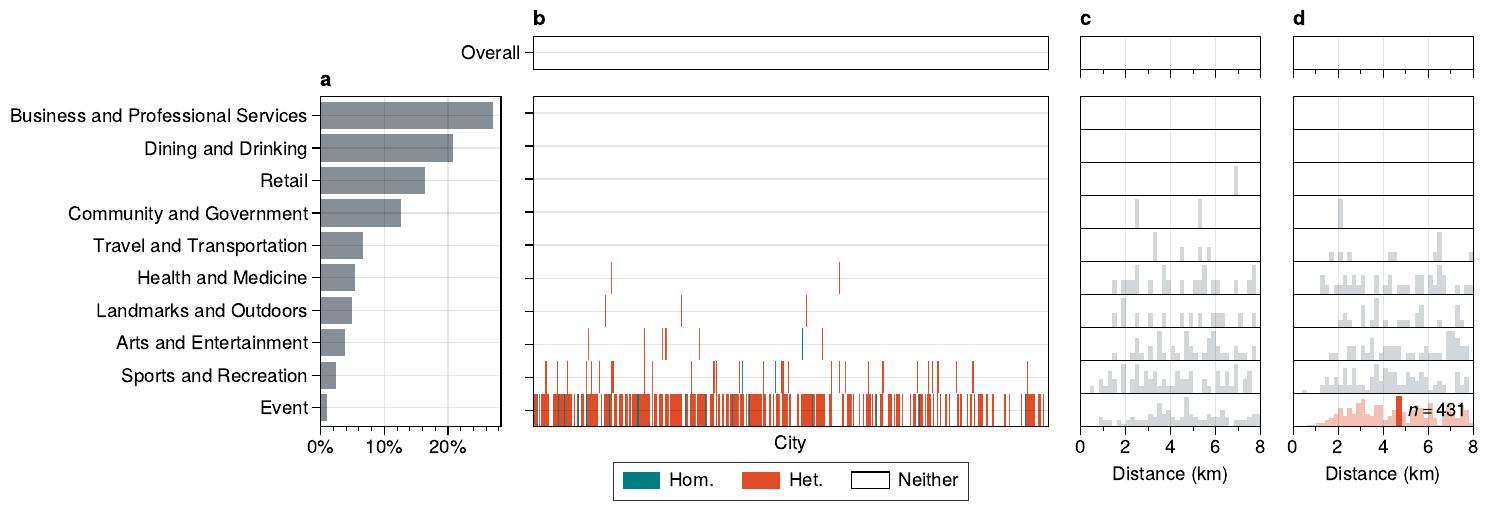}
	\caption{
		\textbf{Homophily and heterophily under randomized null model.} We repeat Figure~\ref{fig:results_foursquare}, this time randomly shuffling the amenity labels $\bm{b}$ within each city to destroy the mixing structure while preserving amenity frequencies (panel a). Five repetitions of label randomization were averaged to reduce spurious fluctuations. The unstructured model of random mixing is now preferred in nearly all cases (panel b), and the characteristic spatial scales of homophily (c) and heterophily (d) are now dispersed across distance scales. These results confirm that the results in Fig.~\ref{fig:results_foursquare} are driven by the precise spatial organization of amenities rather than category abundances. The `Event' amenity class is an outlier, displaying a heterophilic pattern in many cities due to its high level of spatial sparsity resulting in co-agglomeration with other amenity types.
	}
	\label{fig:foursquare_shuffle}
\end{figure}

\begin{figure}[H]
	\centering
	\includegraphics[width=\linewidth]{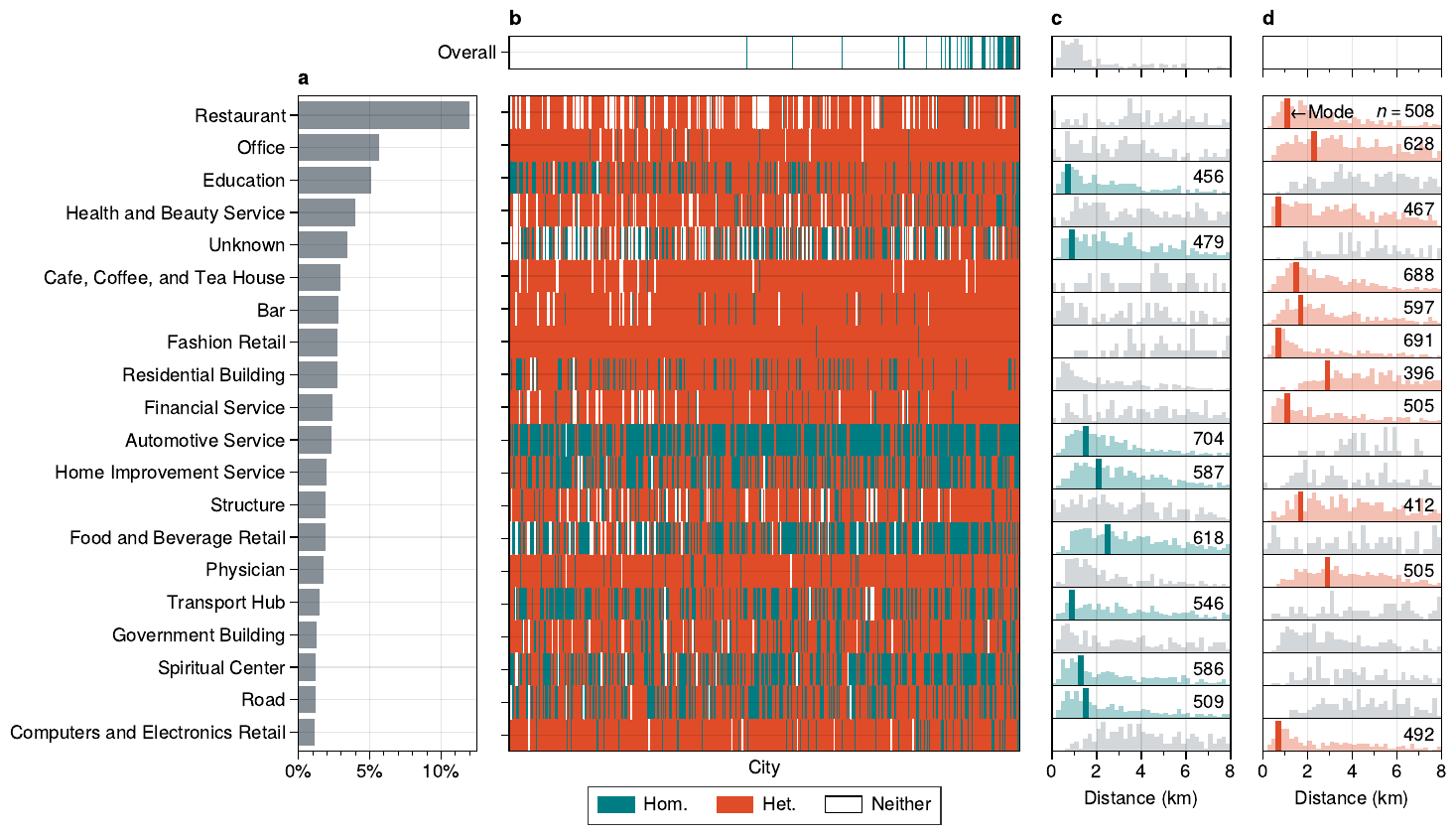}
	\caption{
		\textbf{Homophily and heterophily under fine-grained amenity classification.} We again repeat the analyses of Figure~\ref{fig:results_foursquare}, this time using the 431 fine-grained level-2 amenity categories provided by Foursquare instead of the 10 top-level categories \cite{foursquare}. Only the 20 most frequent level-2 categories (constituting 59.2\% of all POIs) are shown, for easier visualization. We can see that the universal agglomeration preferences and spatial scales identified in Fig.~\ref{fig:results_foursquare} are magnified at this higher resolution, with the vast majority of amenities of the same type having the same homophilic/heterophilic preferences and characteristic spatial scales at which these effects dominate. The `Unknown' category represents amenities that were classified at level-1 but were missing a label for level-2.
	}
	\label{fig:foursquare_lv2}
\end{figure}
\endgroup

\clearpage
\section{Supplementary Case Study Analyses}
\label{sec:supp-casestudy}

\begingroup
\setlength{\intextsep}{0.35\baselineskip}
\setlength{\abovecaptionskip}{0.25\baselineskip}
\setlength{\belowcaptionskip}{0pt}

\begin{figure}[H]
	\centering
	\includegraphics[width=0.85\linewidth]{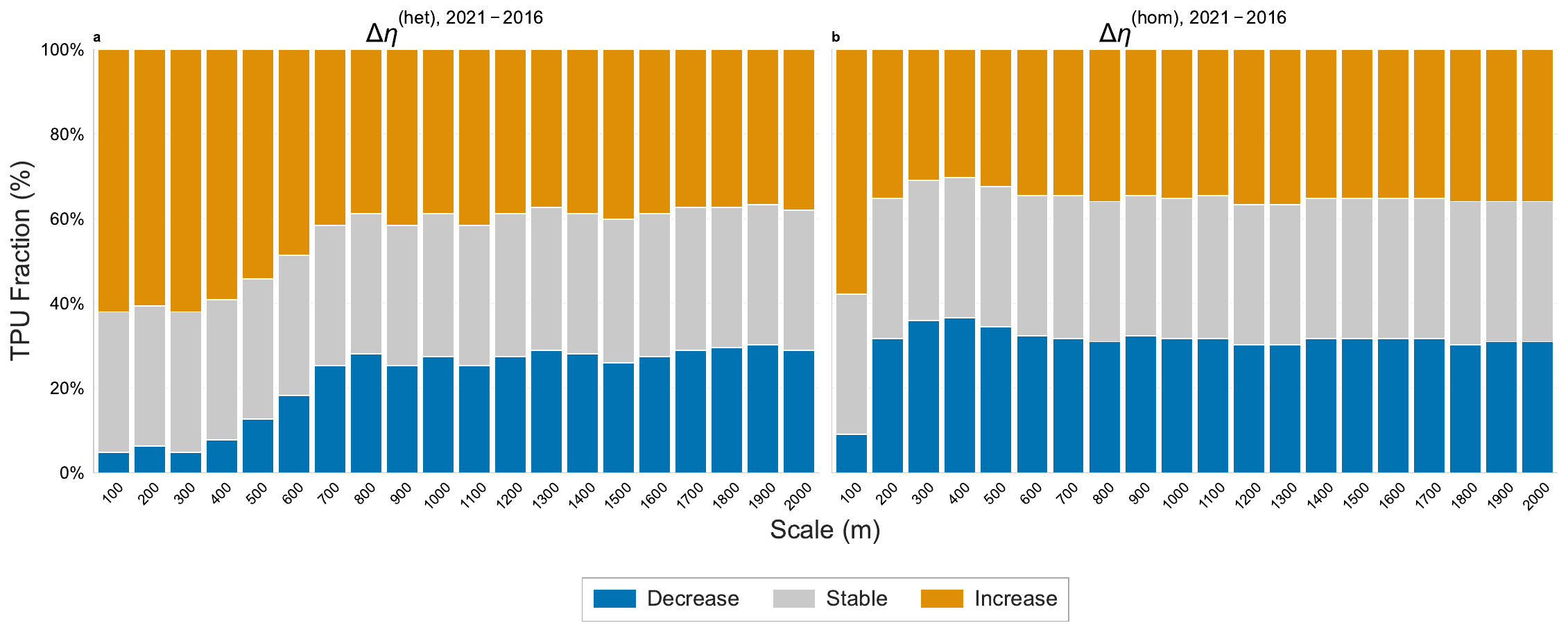}
	\caption{
		\textbf{Sensitivity of compression ratio changes across spatial scales.}
		(a) Fraction of the 142 Hong Kong LTPUGs classified as decreasing, stable, and increasing with respect to their relative change in heterophilic compression ratio (Eq.~\ref{eq:ratio}, see Fig.~\ref{fig:trivariate_het}) from 2016 to 2021, evaluated over neighborhood thresholds $\epsilon=100$--$2000$m.
		(b) Same analyses for the homophilic compression ratio.
		Both results indicate stable mixing profiles across the city at scales of $\sim 800$m and above, ensuring the robustness of the regression analyses in Table~\ref{tab:hk_restaurant_table} to the choice of $900$m scale.  
	}
	\label{fig:cr_across_scales}
\end{figure}

\begin{figure}[H]
	\centering
	\includegraphics[width=0.95\linewidth]{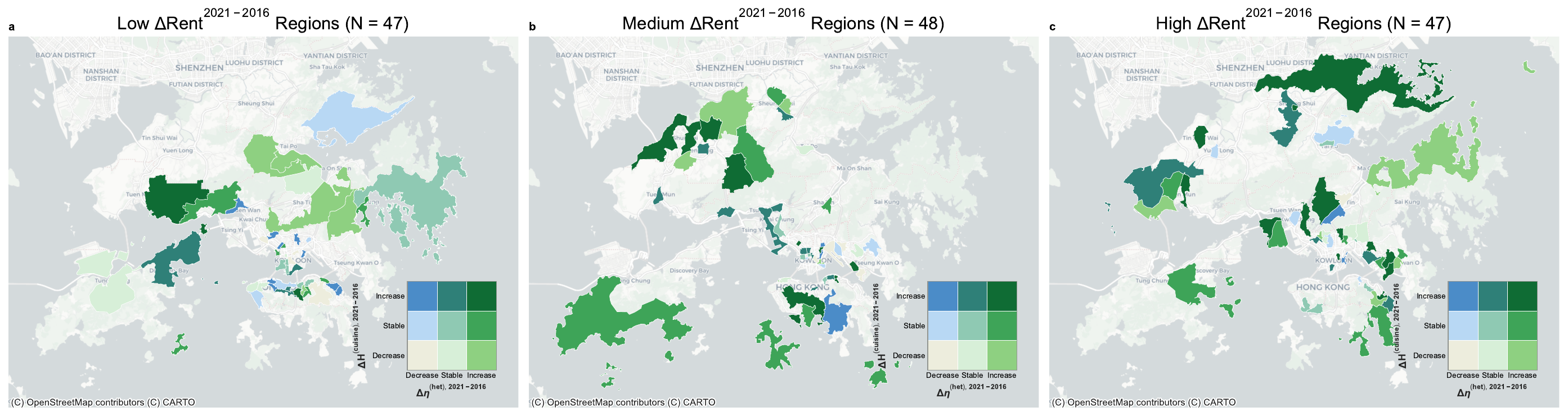}
	\caption{
		\textbf{Trivariate map of cuisine diversity, heterophily compression ratio, and rent change.}
		Hong Kong LTPUGs at $\epsilon=900\mathrm{m}$ partitioned according to relative change in median rent from 2016 to 2021:
		(a) low-rent-change LTPUGs ($N=47$),
		(b) medium-rent-change LTPUGs ($N=48$), and
		(c) high-rent-change LTPUGs ($N=47$).
		Within each panel, colors encode the joint distribution of the relative change in cuisine diversity (Eq.~\ref{eq:entropy}) and relative change in the heterophilic compression ratio (Eq.~\ref{eq:ratio}), with greener cells corresponding to jointly increasing diversity and heterophily and bluer cells indicating joint decreases in these quantities. 
	}
	\label{fig:trivariate_het}
\end{figure}

\begin{figure}[H]
	\centering
	\includegraphics[width=0.95\linewidth]{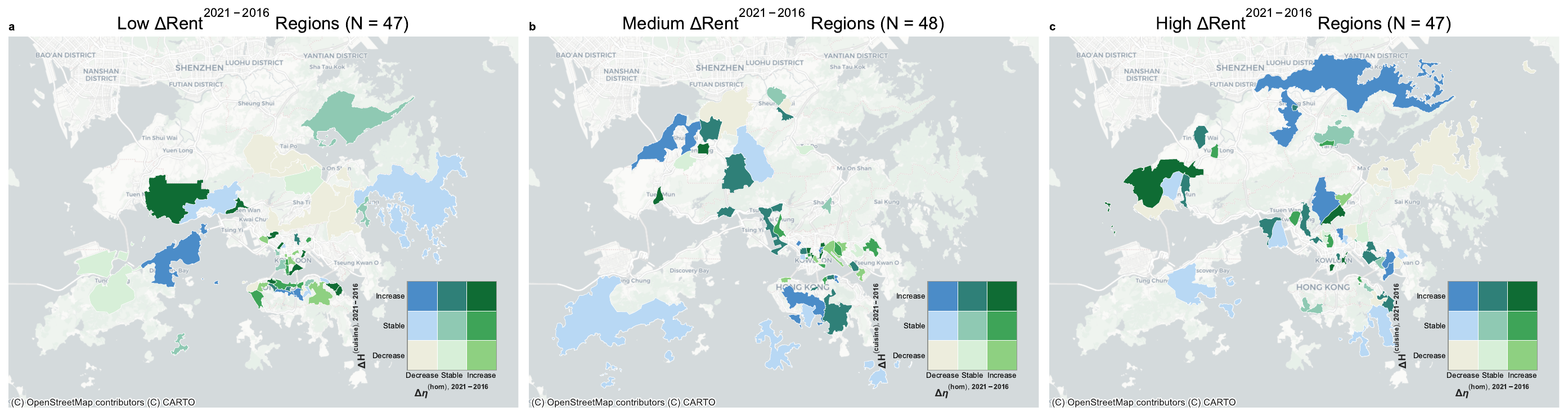}
	\caption{
		\textbf{Trivariate map of cuisine diversity, homophily compression ratio, and rent change.}
		Same rent-stratified classification as Figure~\ref{fig:trivariate_het}, but replacing the heterophilic compression ratio with the homophilic compression ratio. Here we can see, in contrast with Fig.~\ref{fig:trivariate_het}, that high rent growth LTPUGs are not concentrated in cells with jointly increasing diversity and increasing homophily. Instead they are spread across stable and declining homophily classes, including many LTPUGs with rising diversity but falling homophily. This weaker trivariate alignment helps explain why changes in homophily provide less robust explanatory power than changes in heterophily once both are considered together.
	}
	\label{fig:trivariate_hom}
\end{figure}
\endgroup

\clearpage
\section{Foursquare Amenity Data Details}
\label{sec:supp-foursquare-data}

Here we provide the level-2 POIs associated with each of the 10 top-level POIs in the Foursquare dataset used in Sec.~\ref{sec:results}. This gives a more fine-grained look into what types of amenities are present in each category.

\begingroup
\setlength{\intextsep}{0.35\baselineskip}
\setlength{\abovecaptionskip}{0.25\baselineskip}
\setlength{\belowcaptionskip}{0pt}

\begin{itemize}
    \item \textbf{Arts and Entertainment:} Amusement Park, Aquarium, Arcade, Art Gallery, Bingo Center, Bowling Alley, Carnival, Casino, Circus, Comedy Club, Country Club, Country Dance Club, Dance Hall, Disc Golf, Disc Golf Course, Escape Room, Exhibit, Fair, Gaming Cafe, General Entertainment, Go Kart Track, Internet Cafe, Karaoke Box, Laser Tag Center, Mini Golf Course, Movie Theater, Museum, Night Club, Pachinko Parlor, Party Center, Performing Arts Venue, Planetarium, Pool Hall, Psychic and Astrologer, Public Art, Roller Rink, Salsa Club, Samba School, Stadium, Strip Club, Ticket Seller, VR Cafe, Water Park, Zoo.
    \item \textbf{Business and Professional Services:} Advertising Agency, Agriculture and Forestry Service, Appraiser, Architecture Firm, Art Restoration Service, Art Studio, Audiovisual Service, Auditorium, Automation and Control System, Automotive Service, Ballroom, Business Center, Business Service, Career Counselor, Chemicals and Gasses Manufacturer, Child Care Service, Computer Repair Service, Construction, Convention Center, Creative Service, Design Studio, Direct Mail and Email Marketing Service, Distribution Center, Electrical Equipment Supplier, Employment Agency, Engineer, Entertainment Agency, Entertainment Service, Equipment Rental Service, Event Service, Event Space, Factory, Film Studio, Financial Service, Food and Beverage Service, Funeral Home, Geological Service, Health and Beauty Service, Home Improvement Service, Human Resources Agency, Import and Export Service, Industrial Equipment Supplier, Industrial Estate, Insurance Agency, Laboratory, Laundromat, Laundry Service, Leather Supplier, Legal Service, Locksmith, Logging Service, Lottery Retailer, Machine Shop, Management Consultant, Manufacturer, Market Research and Consulting Service, Media Agency, Metals Supplier, Mobile Company, Office, Online Advertising Service, Outdoor Event Space, Paper Supplier, Pet Service, Petroleum Supplier, Photography Service, Plastics Supplier, Power Plant, Print, TV, Radio and Outdoor Advertising Service, Promotional Item Service, Public Relations Firm, Publisher, Radio Station, Real Estate Service, Recording Studio, Recycling Facility, Refrigeration and Ice Supplier, Renewable Energy Service, Rental Service, Repair Service, Research Laboratory, Research Station, Rubber Supplier, Salvage Yard, Scientific Equipment Supplier, Search Engine Marketing and Optimization Service, Security and Safety, Shipping, Freight, and Material Transportation Service, Shoe Repair Service, Storage Facility, TV Station, Tailor, Technology Business, Telecommunication Service, Translation Service, Tutoring Service, Warehouse, Waste Management Service, Water Treatment Service, Wedding Hall, Welding Service, Wholesaler, Writing, Copywriting and Technical Writing Service.
    \item \textbf{Community and Government:} Addiction Treatment Center, Animal Shelter, Assisted Living, Cemetery, Community Center, Cultural Center, Disabled Persons Service, Domestic Abuse Treatment Center, Dump, Education, Government Building, Government Lobbyist, Homeless Shelter, Housing Authority, Housing Development, Library, Observatory, Organization, Polling Place, Prison, Public Bathroom, Public and Social Service, Rehabilitation Center, Residential Building, Retirement Home, Senior Citizen Service, Social Club, Spiritual Center, Summer Camp, Town Hall, Trailer Park, Utility Company.
    \item \textbf{Dining and Drinking:} Bagel Shop, Bakery, Bar, Breakfast Spot, Brewery, Cafe, Coffee, and Tea House, Cafeteria, Cidery, Creperie, Dessert Shop, Distillery, Donut Shop, Food Court, Food Stand, Food Truck, Juice Bar, Meadery, Night Market, Restaurant, Smoothie Shop, Snack Place, Vineyard, Winery.
    \item \textbf{Event:} Conference, Convention, Entertainment Event, Line, Marketplace, Other Event.
    \item \textbf{Health and Medicine:} AIDS Resource, Acupuncture Clinic, Alternative Medicine Clinic, Assisted Living Service, Blood Bank, Chiropractor, Dentist, Emergency Service, Healthcare Clinic, Home Health Care Service, Hospice, Hospital, Maternity Clinic, Medical Center, Medical Lab, Mental Health Service, Nurse, Nursing Home, Nutritionist, Optometrist, Other Healthcare Professional, Physical Therapy Clinic, Physician, Podiatrist, Sports Medicine Clinic, Urgent Care Center, Veterinarian, Weight Loss Center, Women's Health Clinic.
    \item \textbf{Landmarks and Outdoors:} Bathing Area, Bay, Beach, Bike Trail, Boat Launch, Botanical Garden, Bridge, Campground, Canal, Canal Lock, Castle, Cave, Dam, Dive Spot, Farm, Field, Forest, Fountain, Garden, Harbor or Marina, Hiking Trail, Hill, Historic and Protected Site, Hot Spring, Island, Lake, Lighthouse, Memorial Site, Monument, Mountain, Mountain Hut, Nature Preserve, Nudist Beach, Other Great Outdoors, Palace, Park, Pedestrian Plaza, Picnic Shelter, Plaza, Reservoir, River, Rock Climbing Spot, Roof Deck, Scenic Lookout, Sculpture Garden, Stable, States and Municipalities, Structure, Surf Spot, Tree, Tunnel, Volcano, Waterfall, Waterfront, Well, Windmill.
    \item \textbf{Retail:} Adult Store, Antique Store, Arts and Crafts Store, Auction House, Auto Workshop, Automotive Retail, Automotive Shop, Baby Store, Betting Shop, Big Box Store, Board Store, Bookstore, Boutique, Cannabis Store, Comic Book Store, Computers and Electronics Retail, Construction Supplies Store, Convenience Store, Cosmetics Store, Costume Store, Dance Store, Department Store, Discount Store, Drugstore, Duty-free Store, Eyecare Store, Fashion Retail, Financial or Legal Service, Fireworks Store, Flea Market, Floating Market, Flower Store, Food and Beverage Retail, Framing Store, Furniture and Home Store, Garden Center, Gift Store, Hardware Store, Hobby Store, Knitting Store, Leather Goods Store, Luggage Store, Marijuana Dispensary, Market, Medical Supply Store, Miscellaneous Store, Mobility Store, Music Store, Newsagent, Newsstand, Office Supply Store, Outdoor Supply Store, Outlet Mall, Outlet Store, Packaging Supply Store, Party Supply Store, Pawn Shop, Perfume Store, Pet Supplies Store, Pharmacy, Pop-Up Store, Print Store, Record Store, Shopping Mall, Shopping Plaza, Smoke Shop, Souvenir Store, Sporting Goods Retail, Stationery Store, Supplement Store, Swimming Pool Supply Store, Textiles Store, Tobacco Store, Toy Store, Vape Store, Video Store, Vintage and Thrift Store, Warehouse or Wholesale Store.
    \item \textbf{Sports and Recreation:} Athletic Field, Baseball, Basketball, Bowling Green, Cricket Ground, Curling Ice, Equestrian Facility, Fishing Area, Football, Golf, Gun Range, Gym and Studio, Gymnastics, Hockey, Hunting Area, Indoor Play Area, Martial Arts Dojo, Paintball Field, Personal Trainer, Race Track, Racquet Sports, Recreation Center, Rugby, Running and Track, Sauna, Skating, Skydiving Center, Snow Sports, Soccer, Sports Club, Volleyball Court, Water Sports.
    \item \textbf{Travel and Transportation:} Baggage Locker, Bike Rental, Boat Rental, Boat or Ferry, Border Crossing, Cable Car, Cruise, Electric Vehicle Charging Station, Fuel Station, General Travel, Hot Air Balloon Tour Agency, Lodging, Moving Target, Parking, Pier, Platform, Port, RV Park, Rest Area, Road, Street, Toll Booth, Toll Plaza, Tourist Information and Service, Train, Transport Hub, Transportation Service, Travel Agency, Travel Lounge, Truck Stop.
\end{itemize}

\clearpage
\begingroup
\renewcommand{\arraystretch}{0.68}
\setlength{\tabcolsep}{5pt}
\setlength{\LTcapwidth}{0.95\textwidth}
\begin{longtable}{>{\tiny}l>{\tiny}r>{\tiny}r>{\tiny}r>{\tiny}r>{\tiny}r>{\tiny}r>{\tiny}r>{\tiny}r>{\tiny}r>{\tiny}r>{\tiny}r}
\caption{\textbf{Foursquare POI category distributions for 779 cities.} Each row reports the total number of POIs and the percentage share of each of the 10 top-level categories. Only cities with at least 10{,}000 POIs are included.} \label{tab:foursquare} \\
\hline
\multicolumn{1}{c}{\tiny \textsf{\textbf{City}}} & \multicolumn{1}{c}{\tiny \textsf{\textbf{Total}}} & \multicolumn{1}{c}{\tiny \textsf{\textbf{\makecell{Event}}}} & \multicolumn{1}{c}{\tiny \textsf{\textbf{\makecell{Sports and\\Recreation}}}} & \multicolumn{1}{c}{\tiny \textsf{\textbf{\makecell{Arts and\\Entertainment}}}} & \multicolumn{1}{c}{\tiny \textsf{\textbf{\makecell{Landmarks\\and Outdoors}}}} & \multicolumn{1}{c}{\tiny \textsf{\textbf{\makecell{Health and\\Medicine}}}} & \multicolumn{1}{c}{\tiny \textsf{\textbf{\makecell{Travel and\\Transportation}}}} & \multicolumn{1}{c}{\tiny \textsf{\textbf{\makecell{Community\\and\\Government}}}} & \multicolumn{1}{c}{\tiny \textsf{\textbf{\makecell{Retail}}}} & \multicolumn{1}{c}{\tiny \textsf{\textbf{\makecell{Dining and\\Drinking}}}} & \multicolumn{1}{c}{\tiny \textsf{\textbf{\makecell{Business and\\Professional\\Services}}}} \\
\hline
\endfirsthead
\multicolumn{12}{l}{\tiny \textit{(continued)}} \\
\hline
\multicolumn{1}{c}{\tiny \textsf{\textbf{City}}} & \multicolumn{1}{c}{\tiny \textsf{\textbf{Total}}} & \multicolumn{1}{c}{\tiny \textsf{\textbf{\makecell{Event}}}} & \multicolumn{1}{c}{\tiny \textsf{\textbf{\makecell{Sports and\\Recreation}}}} & \multicolumn{1}{c}{\tiny \textsf{\textbf{\makecell{Arts and\\Entertainment}}}} & \multicolumn{1}{c}{\tiny \textsf{\textbf{\makecell{Landmarks\\and Outdoors}}}} & \multicolumn{1}{c}{\tiny \textsf{\textbf{\makecell{Health and\\Medicine}}}} & \multicolumn{1}{c}{\tiny \textsf{\textbf{\makecell{Travel and\\Transportation}}}} & \multicolumn{1}{c}{\tiny \textsf{\textbf{\makecell{Community\\and\\Government}}}} & \multicolumn{1}{c}{\tiny \textsf{\textbf{\makecell{Retail}}}} & \multicolumn{1}{c}{\tiny \textsf{\textbf{\makecell{Dining and\\Drinking}}}} & \multicolumn{1}{c}{\tiny \textsf{\textbf{\makecell{Business and\\Professional\\Services}}}} \\
\hline
\endhead
\hline
\endlastfoot
\makecell[l]{A Coruna, ES} & 13852 & 0.25\% & 1.86\% & 3.08\% & 4.15\% & 5.05\% & 5.50\% & 7.30\% & 23.61\% & 27.45\% & 21.75\% \\
\makecell[l]{Aachen, DE} & 14396 & 0.58\% & 1.83\% & 2.45\% & 3.11\% & 9.57\% & 6.42\% & 10.86\% & 18.12\% & 12.25\% & 34.82\% \\
\makecell[l]{Aberdeen, GB} & 10305 & 0.21\% & 1.91\% & 2.14\% & 2.27\% & 4.34\% & 7.28\% & 10.91\% & 19.45\% & 14.43\% & 37.06\% \\
\makecell[l]{Abu Dhabi, AE} & 19695 & 0.57\% & 2.47\% & 1.81\% & 4.89\% & 3.32\% & 6.15\% & 11.62\% & 16.60\% & 30.19\% & 22.38\% \\
\makecell[l]{Acapulco de Ju\'arez, MX} & 12005 & 0.53\% & 3.03\% & 3.44\% & 4.46\% & 5.64\% & 7.17\% & 12.34\% & 14.85\% & 28.85\% & 19.68\% \\
\makecell[l]{Ad Dammam, SA} & 16758 & 1.10\% & 1.52\% & 1.34\% & 5.29\% & 3.50\% & 6.43\% & 13.27\% & 15.70\% & 28.22\% & 23.63\% \\
\makecell[l]{Ad Dawhah, QA} & 23273 & 0.69\% & 2.77\% & 2.42\% & 5.19\% & 3.57\% & 8.90\% & 12.42\% & 15.36\% & 27.83\% & 20.85\% \\
\makecell[l]{Adana, TR} & 120047 & 2.42\% & 1.59\% & 6.40\% & 5.97\% & 2.37\% & 5.01\% & 18.61\% & 16.10\% & 16.80\% & 24.73\% \\
\end{longtable}
\endgroup

\endgroup

\end{document}